\documentclass[12pt,american,english]{article}
\usepackage[LGR,T1]{fontenc}
\usepackage[latin9]{inputenc}
\usepackage{geometry}
\usepackage{graphicx}
\usepackage{setspace}
\makeatletter

\DeclareRobustCommand{\greektext}{%
  \fontencoding{LGR}\selectfont\def\encodingdefault{LGR}}
\DeclareRobustCommand{\textgreek}[1]{\leavevmode{\greektext #1}}
\ProvideTextCommand{\~}{LGR}[1]{\char126#1}

\newcommand{\lyxaddress}[1]{
\par {\raggedright #1
\vspace{1.4em}
\noindent\par}
}

\@ifundefined{date}{}{\date{}}
\usepackage{lineno}
\usepackage{floatpag}

\floatpagestyle{empty} 

\makeatother

\usepackage{babel}
\begin{document}

\title{A Multiple Impact Origin for the Moon}
\maketitle
\begin{center}
Raluca Rufu$^{1*}$, Oded Aharonson$^{1}$, Hagai B. Perets$^{2}$
\par\end{center}

\lyxaddress{$\!$$^{1}$Weizmann Institute of Science, Department of Earth and
Planetary Sciences, Rehovot 76100, Israel.}

\lyxaddress{$^{2}$Technion Israel Institute of Technology, Physics Department,
Haifa 32000, Israel.}

\date{$^{*}$Contact Information: e-mail: Raluca.rufu@weizmann.ac.il }

\newpage{}

\textbf{The hypothesis of lunar origin by a single giant impact can
explain some aspects of the Earth-Moon system. However, it is difficult
to reconcile giant impact models with the compositional similarity
of the Earth and Moon without violating angular momentum constraints.
Furthermore, successful giant impact scenarios require very specific
conditions such that they have a low probability of occurring. Here
we present numerical simulations suggesting that the Moon could instead
be the product of a succession of a variety of smaller collisions.
In this scenario, each collision forms a debris disk around the proto-Earth
that then accretes to form a moonlet. The moonlets tidally advance
outward, and may coalesce to form the Moon. We find that sub-lunar
moonlets are a common result of impacts expected onto the proto-Earth
in the early solar system and find that the planetary rotation is
limited by impact angular momentum drain. We conclude that, assuming
efficient merger of moonlets, a multiple impact scenario can account
for the formation of the Earth-Moon system with its present properties.}

The Moon's origin remains enigmatic. The leading theory for the Moon\textquoteright s
formation posits a scenario in which a Mars-sized planetesimal impacts
the late-stage accreting Earth. The ejected material produces an Earth-orbiting
disk, which later gravitationally accretes to a single Moon. Impact
simulations found that the projectile contributes more than 70\% to
the disk\textquoteright s mass\cite{canup2004simulations}. This skewed
mass contribution is a well recognized problem as more high-precision
measurements of isotopes indicate that the Moon and Earth are isotopically
similar in oxygen ($^{17}{\rm O/^{16}O}$ and $^{18}{\rm O}/{}^{16}{\rm O}$
within $12\pm3{\rm \ ppm}$\cite{herwartz2014identification}), titanium
($^{50}{\rm Ti/^{47}Ti}$ within $\pm4{\rm \ ppm}$\cite{zhang2012proto})
and pre-late veneer tungsten ($^{182}W/^{184}W$\cite{Kruijer:2015aa}).
Isotopic equilibration with a hot protoplanetary atmosphere is efficient
for oxygen but insufficient to explain the similarity in more refractory
elements such as titanium\cite{Pahlevan:2007aa,zhang2012proto}. 

Deriving more disk material from the proto-Earth occurs in impact
scenarios with increased angular momentum\cite{Canup23112012,cuk2012making}
beyond the present value, that is later dissipated by an orbital resonance
or an associated limit cycle\cite{Wisdom2015138}. Studies of planetary
accretion\cite{Jacobson20130174,Kaib2015161} have shown that the
equal sized impactors are extremely rare unless assuming a very early
event, which is inconsistent with the recent Moon formation timing
estimates\cite{Jacobson:2014aa}. The expected impacts include Moon
to Mars-sized impactors, supporting the specific suggested scenario
of a fast spinning Earth\cite{cuk2012making}, however the predicted
velocity and impact parameters phase space is much wider than the
preferred range (fast and low angle impactors). Alternatively, N-body
simulations have been used to argue that Earth-impactors are likely
to be compositionally similar to Earth, relative to other bodies\cite{Mastrobuono-Battisti:2015aa}. 

In this paper, we consider a multi-impact hypothesis for Moon's formation\cite{Citron,RINGWOOD1989208}.
In this scenario, the proto-Earth experiences a sequence of collisions
by medium to large size bodies ($0.01-0.1M_{\oplus})$ (Fig. 1-a,
d). Small satellites form from the impact-generated disks (1-b, e)
and migrate outward controlled by tidal interactions, faster at first,
and slower as the body retreats away from the proto-Earth (1-c). The
slowing migration causes the satellites to enter their mutual Hill
radii and eventually coalesce to form the final Moon\cite{Jutzi:2011aa}
(1-f). In this fashion, the Moon forms as a consequence of a variety
of multiple impacts in contrast to a more precisely tuned single impact.
A similar scenario using smaller ($0.001-0.01M_{\oplus}$), high velocity,
late accreting impactors was previously suggested\cite{RINGWOOD1989208},
but supporting calculations were not provided. 

N-body simulations of terrestrial planet accretion\cite{Agnor1999219}
show that the final angular momentum of the Earth's system is a result
of several impacts. The largest impactor is not necessary the last
one. That study shows that majority of single collisions with Earth
cannot form the present Moon because the impact angular momentum is
insufficient relative to that of the current value, $L_{{\rm EM}}$.
Similar to the angular momentum of the Earth-Moon system, we argue
that the Moon's mass is also the result of contributions from several
last impactors. 

In the multi-impact scenario, the 'compositional crisis' described
above is mitigated by two effects. \foreignlanguage{american}{First,
since the mass and angular momentum of the present Earth-Moon system
provide constraints on the sum of multiple impacts, rather than a
single impact}, the additional freedom in impact geometries enables
mining more material from Earth than in the conventional scenario.
Second, the oxygen signature distribution of the cumulative sum of
multiple moonlets will have a reduced variance, increasing the probability
of the Earth-Moon similarity compared to that from a single event.

We investigate the formation of moons by multiple impacts, and consider
if Earth's Moon specifically may have been constructed by such a mechanism.
Because the Moon in this scenario is constructed in parts, we can
sample a large phase space of initial conditions. We choose parameters
for the impactor mass ratio $\gamma$, speed $V_{{\rm imp}}$, direction
angle $\beta$ (relative to the line connecting the centers at contact),
and planetary rotation $\omega$ (see Supplementary Table 1 for exact
values), which resemble the characteristics of the last impactors
onto Earth\cite{Agnor1999219}. High velocity impactors are more frequent
for smaller planetesimals because their eccentricities and inclinations
are strongly affected by scattering events and distant interactions.
In comparison with the classic impact scenario\cite{canup2004simulations},
we consider smaller masses and a larger range of angles (from head-on
to near grazing impacts). In comparison with the preferred scenario
in the fast spinning Earth scenario\cite{cuk2012making}, we include
higher impact velocities and more slowly rotating planets.

\begin{figure}
\begin{centering}
\includegraphics[width=1\textwidth]{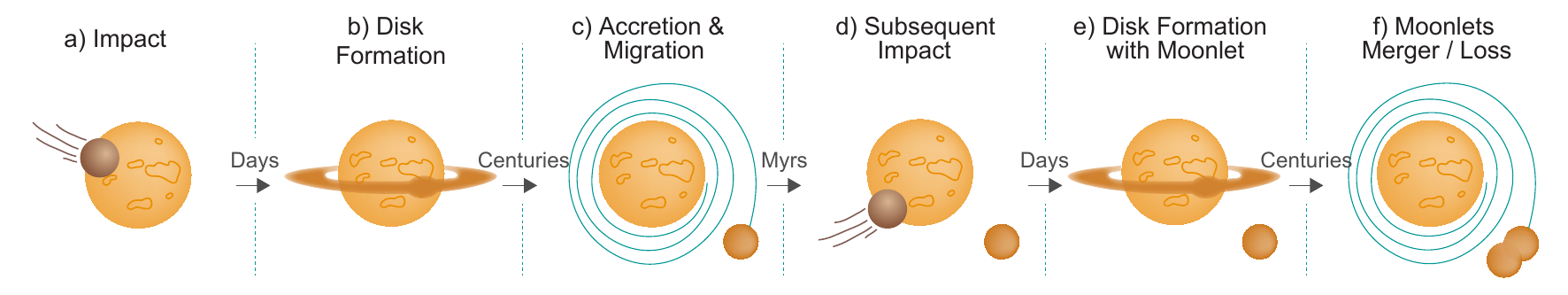}
\par\end{centering}
\caption{\textbf{Lunar formation in the multiple impact scenario.} Moon to
Mars sized bodies impact the proto-Earth (a) forming a debris disk
(b). Due to tidal interaction, accreted moonlets migrate outward (c).
Moonlets reach distant orbits before the next collision (d), and the
subsequent debris disk generation (e). As the moonlet-proto-Earth
distance grows, the tidal acceleration slows, moonlets enter their
mutual Hill radii. The moonlet interactions can eventually lead to
moonlet loss or merger (f). The time scale between these stages is
estimated from previous works \cite{canup2004simulations,ida1997lunar,Raymond2009644}.}
\end{figure}

\begin{figure}
\begin{centering}
\includegraphics[width=1\textwidth]{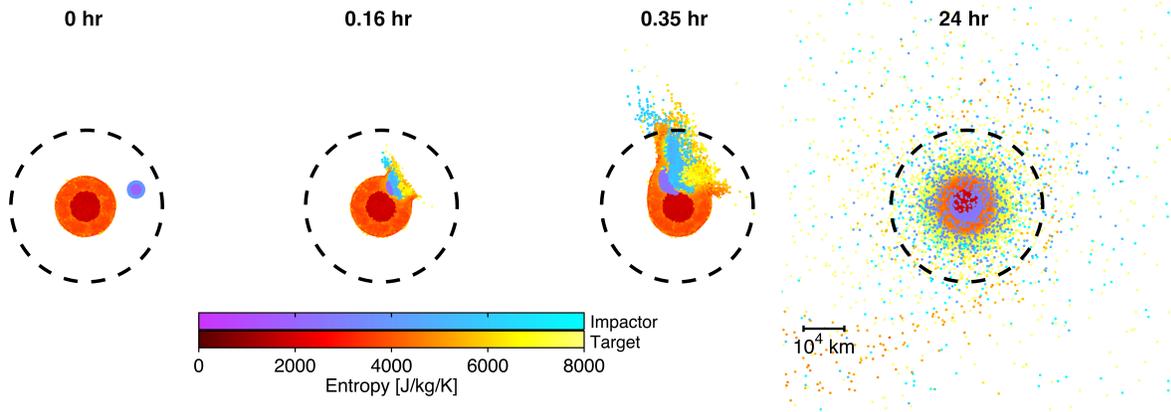}
\par\end{centering}
\caption{\foreignlanguage{american}{\textbf{Impact simulation. }Several snapshots of one of the simulation
with initial conditions of $\gamma=0.025$, $V_{{\rm imp}}=2V_{{\rm esc}}$,
$\beta=30^{o}$ and $\omega=0.5\,\omega_{{\rm max}}$. \foreignlanguage{english}{The
color bars represent the entropy of the impactor and target. All projections
are on the equatorial plane with one hemisphere removed. The impactor
core is shown over the target. The Roche limit is represented by the
dashed line.}}}

\end{figure}

\section*{\label{subsec:Debris-Disk-Mass}Debris Disk }

The mass of the impact-generated debris disks is evaluated relative
to the present Moon's. Of the 864 simulations performed, 750 simulations
result in a disk with a discernible mass at our resolution ($>10^{-3}M_{{\rm moon}}$).
The most massive disk found is $1.20M_{{\rm moon}}$, which results
from a graze-and-merge impact with the highest mass impactor at a
speed of $\sim1V_{{\rm esc}}$, onto the most rapidly rotating target
($\omega=0.5\,\omega_{{\rm max}}$, where $\omega_{{\rm max}}$ is
the planetary rotation rate at tidal break-up) within our parameter
space. 

An off-axis energetic impact onto a rotating target is found to produce
a debris disk that is sourced from both impactor and target material.
The cores are seen to merge, and the heating is concentrated in the
upper planetary mantle (Fig. 2). The high angle impacts ($\beta=45^{o}$)
almost always produce a disk because the angular momentum of the impactor,
even for relatively low velocities near $V_{{\rm esc}},$ is sufficient
to impart enough angular momentum to the ejecta to form a disk (Fig.
3). The head-on and low-angle impacts require a high velocity to eject
material to orbit. Moreover, high-angle (grazing) impacts result in
little mixing between the planet and the disk with the impactor contributing
a substantial fraction to the disk. Low-angle impacts produce disks
with higher degree of compositional similarity to Earth. 

We find the composition of the disk does not strongly depend upon
the retrograde sense ($L_{{\rm imp}}<0$, where the impactor's angular
momentum is opposite to that of the planet) or prograde sense ($L_{{\rm imp}}>0$)
of the impact. 

Several trends in disk mass emerge. The majority of initial conditions
within the wide phase space we tested result in disks that yield a
high compositional difference, $|\delta f_{{\rm T}}|>10\%$. However,
the fast impactors produce disks with a low value of $\delta f_{{\rm T}}$
and a lack of iron. Significant fractions of the mass of fast, high
angle impactors escape the system, leaving material originating from
the planet's mantle in orbit.

In addition to these trends, we also find abrupt transitions in disk
mass. At high impact angles and low velocities, disk masses span a
wide range varying by as much as a factor of ten. These scenarios
transition from graze-and-merge to partial-accretion (previously defined\cite{stewart2012collisions})
as the mass ratio increases. The graze-and-merge scenario exhibits
a qualitatively different character, as the impactor collides with
the target a second time resulting in significantly enhanced mass
ejection. 

\begin{figure}
\begin{raggedright}
\textbf{a)}
\par\end{raggedright}
\begin{centering}
\includegraphics[scale=0.3]{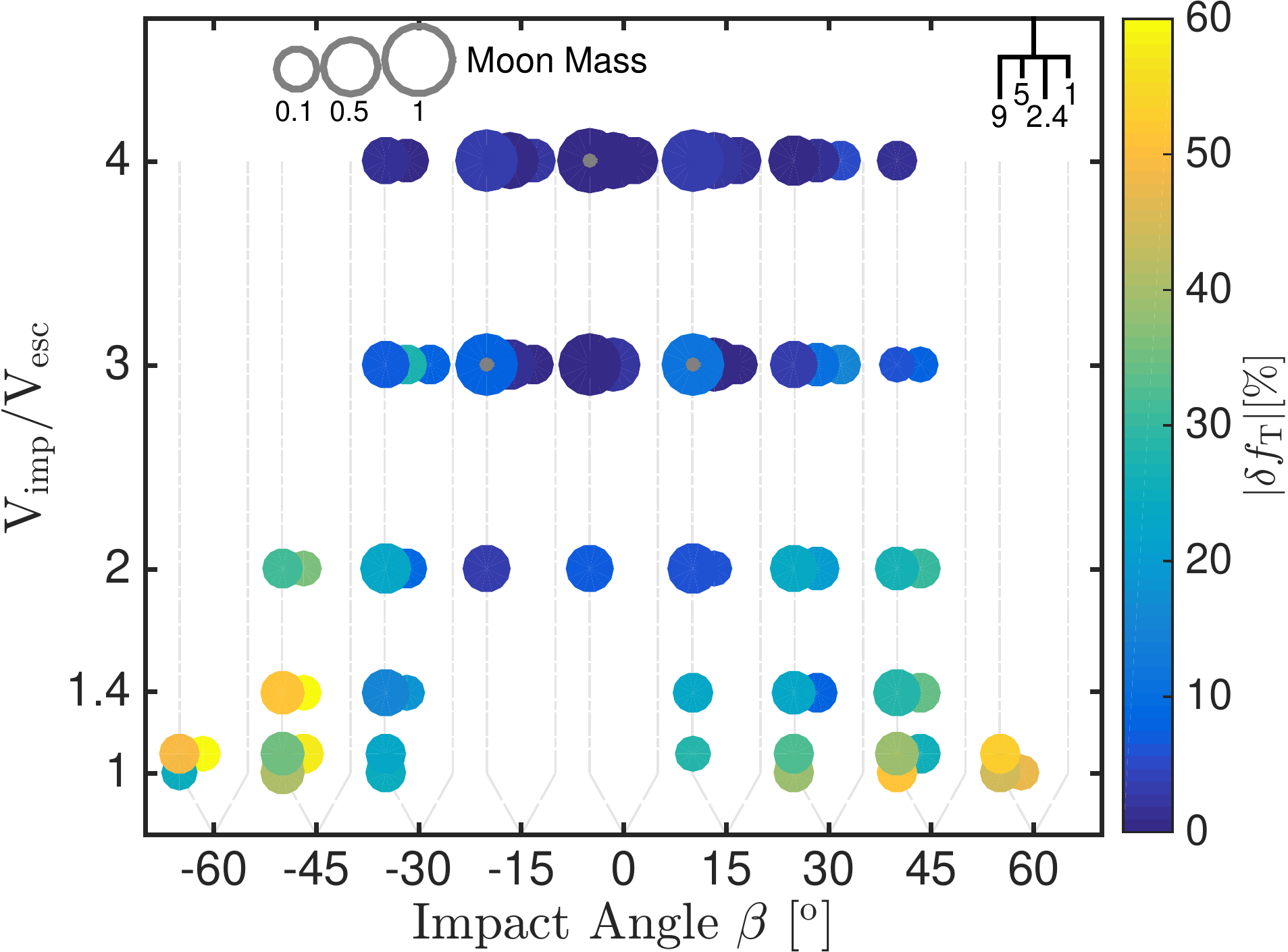} 
\par\end{centering}
\begin{raggedright}
\textbf{b)}
\par\end{raggedright}
\begin{centering}
\includegraphics[scale=0.3]{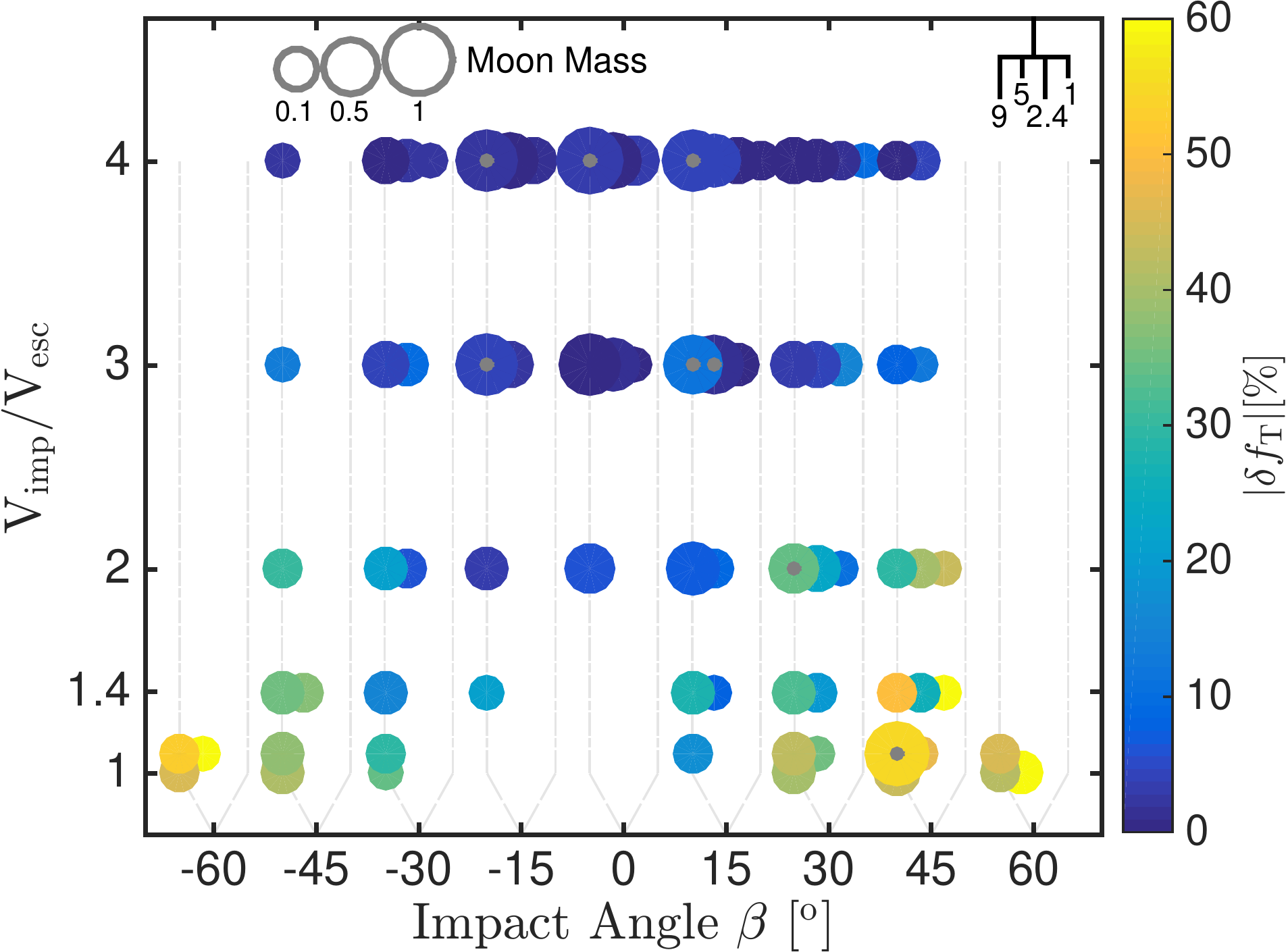}
\par\end{centering}
\begin{raggedright}
\textbf{c)}
\par\end{raggedright}
\begin{centering}
\includegraphics[scale=0.3]{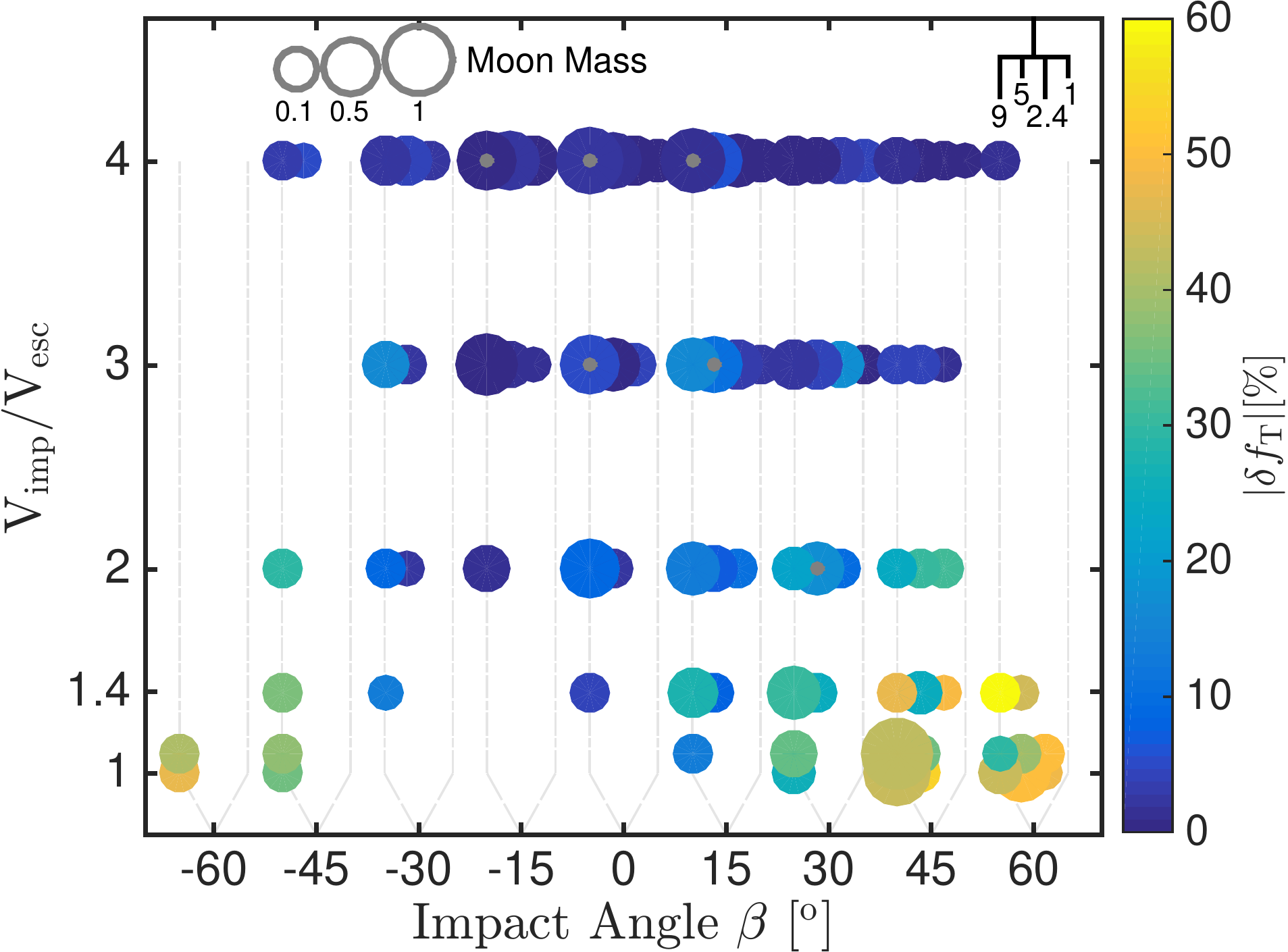}
\par\end{centering}
\caption{\textbf{Disk properties in the angle-velocity phase space.} Initial
planetary rotation rate a) $\omega=0.10\,\omega_{{\rm max}}$, b)
$\omega=0.25\,\omega_{{\rm max}}$, and c) $\omega=0.50\,\omega_{{\rm max}}$.
The marker size corresponds to disk mass and the color to the compositional
difference between the silicates in the final planet and disk. For
comparison, the grey circles in the upper left corner represent a
disk mass of $0.1$, $0.5$ and $1$ $M_{{\rm Moon}}$. Markers are
shifted horizontally according to the mass ratio, from left to right
{[}9, 5, 2.4, 1\%{]}. The grey dot indicates disks that have an iron
content larger than the estimated lunar core mass of $0.1M_{{\rm moon}}$.
Disks containing $<100$ SPH particles were omitted. }
\end{figure}

The mass of the satellite increases with impact angular momentum in
the partial-accretion regime (Fig. 4-a). Again, the maximum satellite
mass is achieved by impacts that graze-and-merge. Satellites formed
by hit and run collision show large relative mass variations but they
are all smaller than $\sim0.2\,M_{{\rm Moon}}$. 

For head-on impacts only planets with initial rotation produce a moonlet
because the angular momentum required for the disk to remain stable
against collapse must originate from the planetary rotation (Fig.
4-b). Head\textendash on impacts with slow rotation produce disks
with low contribution of material from the impactor. However, they
lack angular momentum and a moonlet cannot form. Adding rotation to
the planet increases the angular momentum of the disks, rendering
the initial rotation of the planet crucial in producing moonlets by
low angle impacts. The energetic hit-and-run regimes are easily distinguishable
as the final angular momentum is close to the initial angular momentum
(represented by the dashed lines), independent on the impactor's angular
momentum. In contrast, in the accretionary scenarios the angular momentum
changes by an amount proportional to the impactor's initial angular
momentum. The resulting systems sometimes exceed $L_{{\rm EM}}$ although
the values are lower than those found in past work\cite{cuk2012making,Canup23112012}
because the impacts here are somewhat smaller. We note that subsequent
impacts in a multiple impact scenario may further alter this value.
The high angular momentum and high energy cases can cross the ``hot
spin stability limit''\cite{LockStewart}, explaining the Moon's
enrichment in heavy potassium isotopes\cite{Wang:2016aa}.

\begin{figure}
\begin{raggedright}
\textbf{a)}
\par\end{raggedright}
\begin{centering}
\includegraphics[scale=0.4]{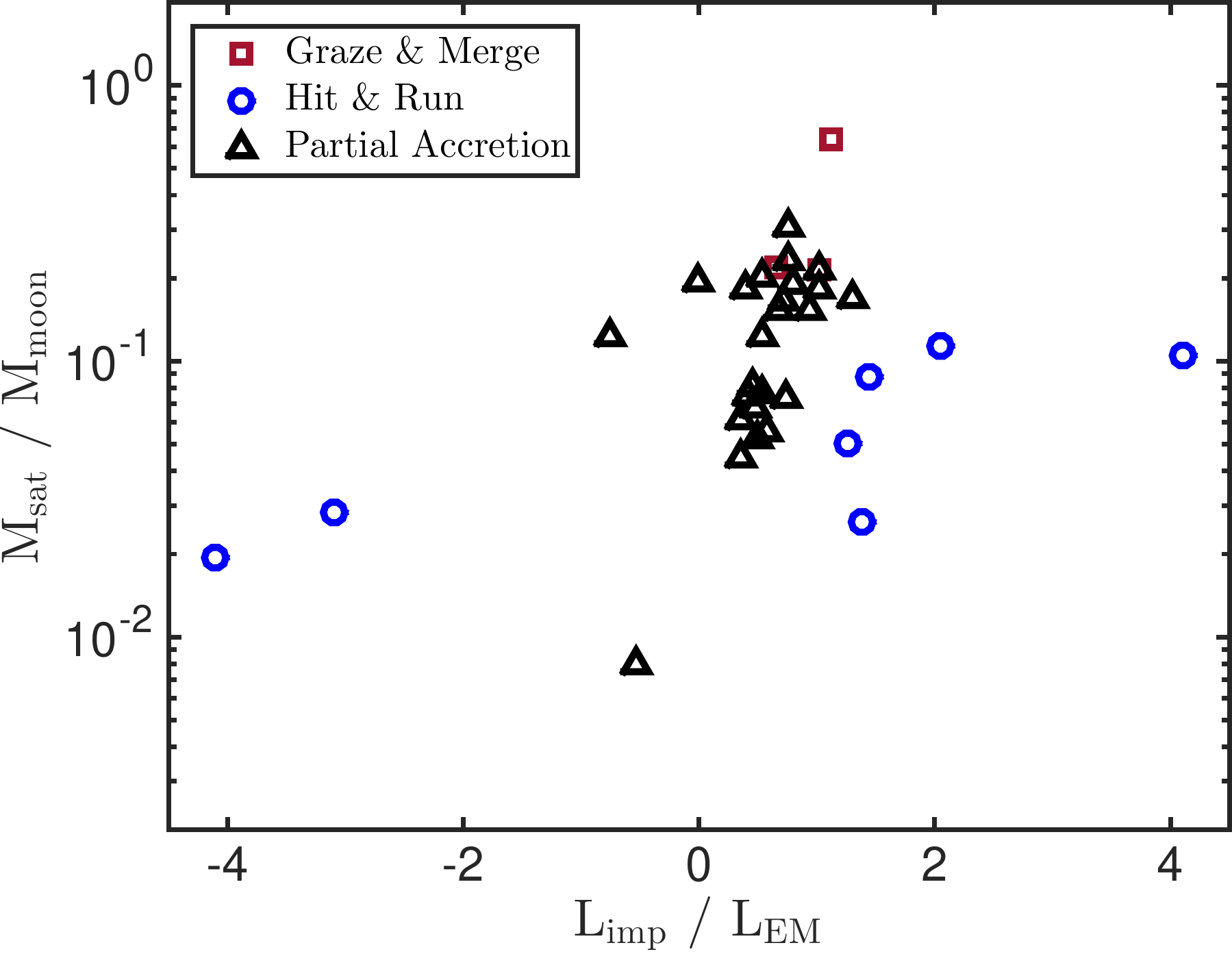}
\par\end{centering}
\begin{raggedright}
\textbf{b)}
\par\end{raggedright}
\begin{centering}
\includegraphics[scale=0.4]{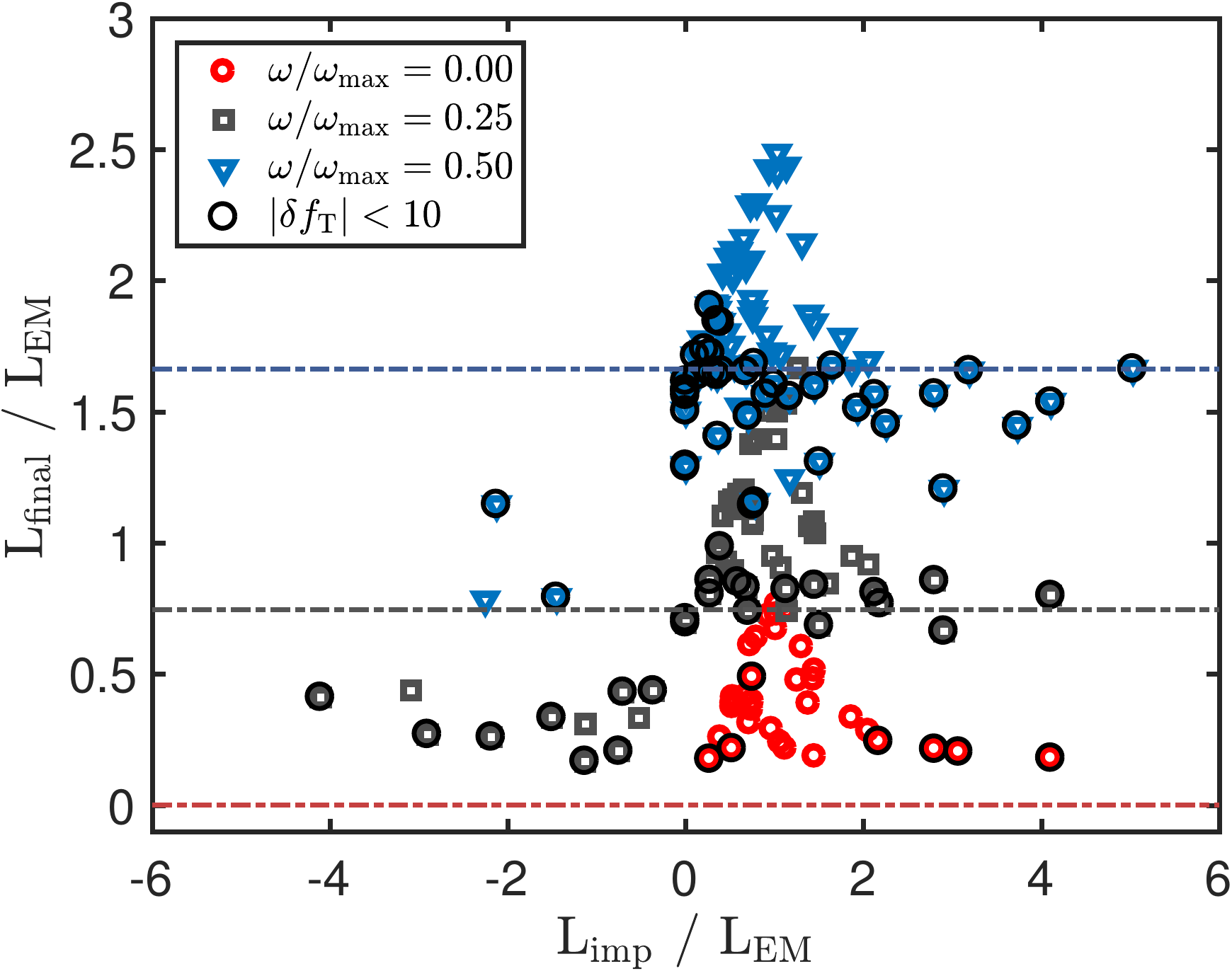}
\par\end{centering}
\caption{\textbf{Final satellite mass and system angular momentum.} a) Mass
of the formed satellite using the equation (\ref{eq:M_sat}) as a
function of impact angular momentum for $\omega=0.25\,\omega_{{\rm max}}$
rotation. Colors correspond to different collisional regimes (Hit
\& Run - impactor escapes partially intact; Graze \& Merge - impactor
impacts the target twice; partial accretion - addition of mass to
the target); b) The final angular momentum of all the systems which
created a satellite. The different style of markers represent different
initial rotations. The darker horizontal lines represent the initial
planetary angular momentum value with colors corresponding to the
colors of the markers. Disks containing $<100\ N_{{\rm SPH}}$ were
omitted.}
\end{figure}

We conclude that medium impactors ($0.01-0.1M_{\oplus}$) can produce
a sub-lunar size moonlet whose composition ranges from impactor dominated
to target dominated. Near head-on impactors are preferred because
they efficiently incorporate planetary material in the disk. Within
the parameter space investigated, we find that the retrograde impactors
often fail to form a disk with enough angular momentum to accrete
a moonlet.

\subsection*{Proto-Earth}

Post impact the planet may gain or lose mass depending on the balance
of impactor merger and net erosion of material. Overall, the majority
of the cases examined produce a partial or perfect merger; high energy
cases result in net erosion of the planet (Supplementary Fig. 3).
Extreme cases ($\gamma=0.091$; $V_{{\rm imp}}=4V_{{\rm esc}}$; $\beta=0^{o}$
at all initial rotations) even strip half the planet's mass. These
findings are in good agreement with past studies\cite{stewart2012collisions},
which found that planetary erosion is more common at lower impact
angles than at high angles due to the larger interacting mass and
interacting energy (defined in the Methods section).

For initially slowly rotating targets, the spin accelerates (the period
decreases) due to typical prograde impacts (Fig. 5). However the rotation
rate saturates for more rapidly rotating planets, which are difficult
to accelerate further due to angular momentum drain carried away by
ejected material. The retrograde impacts consistently decelerate the
planet. The planetary angular momentum may be decreased by both material
ejection and retrograde impacts. Hence, in the limit of a large number
of impacts, angular momentum drain limits the otherwise random-walk
growth of the angular momentum vector. As shown, for an initial rotation
rate of 5.9 hours ($\omega=0.25\,\omega_{{\rm max}}$), retrograde
collisions decelerate the planet while prograde collisions hardly
change the period, so that accelerating beyond this period is difficult. 

\selectlanguage{american}%
Moreover, changes in the direction of the planetary spin in one single
event is seldom large. We observed 3 retrograde cases where the small
planetary spin (\foreignlanguage{english}{$\omega=0.1\,\omega_{{\rm max}}$})
changed its direction. Therefore we conclude that usually subsequent
moonlets will have the same sense of rotation (both prograde or both
retrograde).

\begin{figure}
\selectlanguage{english}%
\begin{centering}
\includegraphics[width=0.5\textwidth]{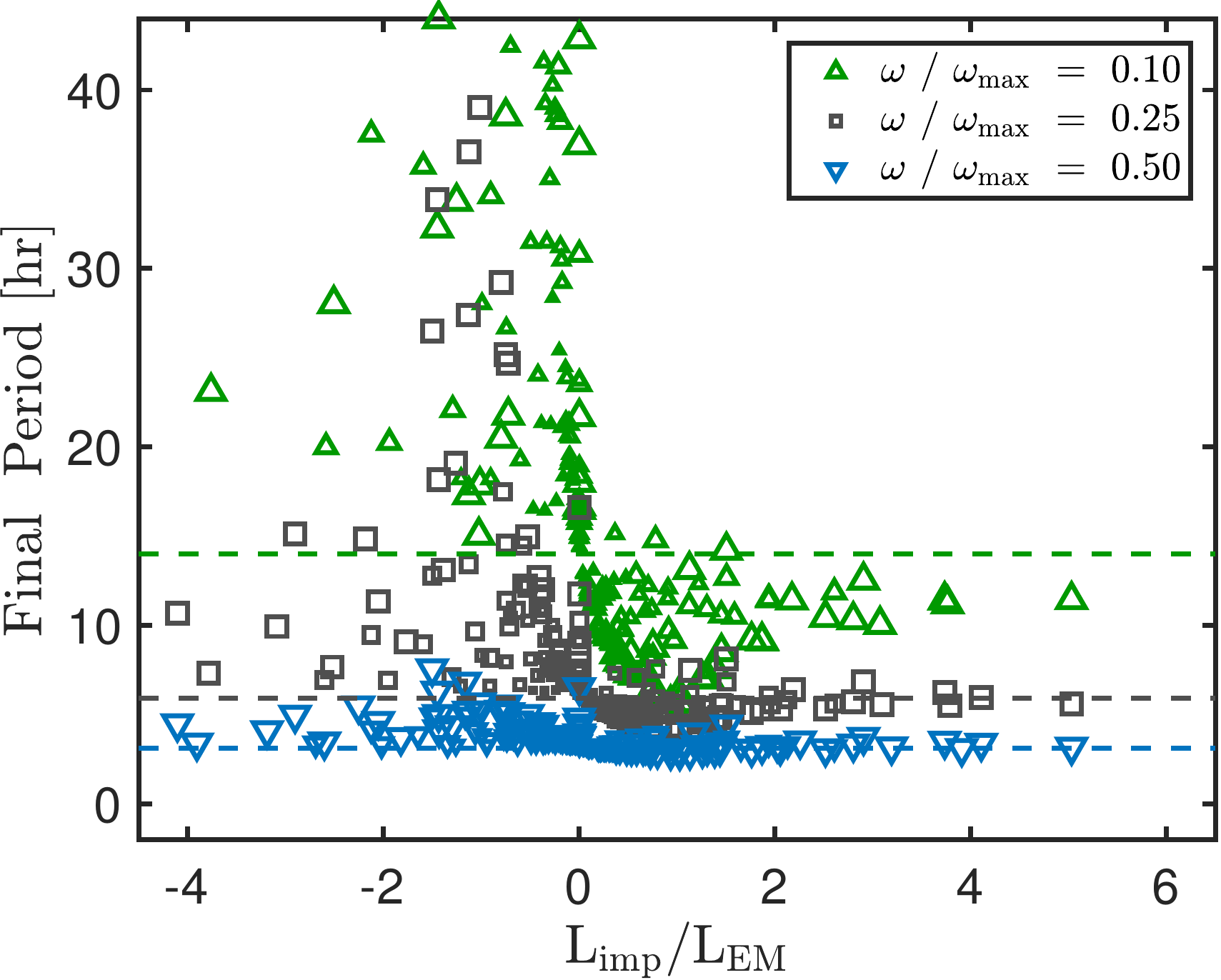}
\par\end{centering}
\selectlanguage{american}%
\caption{\foreignlanguage{english}{\textbf{Planetary rotation}. Post impact rotation as a function
of the initial impact angular momentum. Marker size represents the
mass of the impactor and marker style and color represent the initial
rotation rate of the target, also indicated by the dashed lines for
each group.}}
\end{figure}
\selectlanguage{english}%

\subsection*{Disk Structure}

The disk structure, including its entropy, density and vapor fraction
are important in determining the constituents and efficiency of moonlet
accretion. Tidal forces within the Roche limit prevent accretion,
therefore the inner disk material must spread beyond the Roche limit
before it can aggregate. The accretion of material outside the Roche
limit is efficient, whereas the accretion of the inner disk is self-limiting
because the newly spawned moonlets at the boundary will confine the
disk and cause disk mass to fall onto Earth\cite{Salmon20130256}.
In the accretionary simulations here disks of non-negligible mass
are well extended beyond the Roche limit, with more than 60\% of their
mass outside this radius. Although initial rotation has been shown
to somewhat reduce this fraction\cite{Nakajima2014259}, we conclude
here that the radial mass distribution of the disk is amenable to
moonlet formation, subject to angular momentum constraints (discussed
above).

The surface density is maximum inside the Roche limit, but is lower than
in previously proposed scenarios (\cite{Canup23112012,cuk2012making,canup2004simulations}
as calculated by\cite{Nakajima2014259}) due to the lower disk mass
in our scenario (Fig. 6-a,c). The initial entropy of the the disk
will determine the amount of volatiles incorporated in the disk and
the cooling time required to condense the silicate atmosphere\cite{Pahlevan:2007aa}.
We find the specific entropy is approximately constant throughout
the disk, with deviations of only $\sim$10\% from the mean (Fig.
6-b-d). The mean entropy grows with impact energy, but while the impact
energy ranges over more than two orders of magnitude, the mean entropy
increases by only a factor of three. This may be understood by recognizing
that the shock wave which accelerates material to bound orbits, is
also responsible for heating the material. Thus the addition of kinetic
and thermal energies are linked and ejected material which remains
bound in the disk cannot be heated arbitrarily. When significant erosion
occurs, the disk entropy is reduced because more energy is expended
in ejecting material to escaping trajectories. 

\begin{figure}
\textbf{a)}\hspace{0.29\textwidth}\textbf{c)}\hspace{0.23\textwidth}

\includegraphics[scale=0.28]{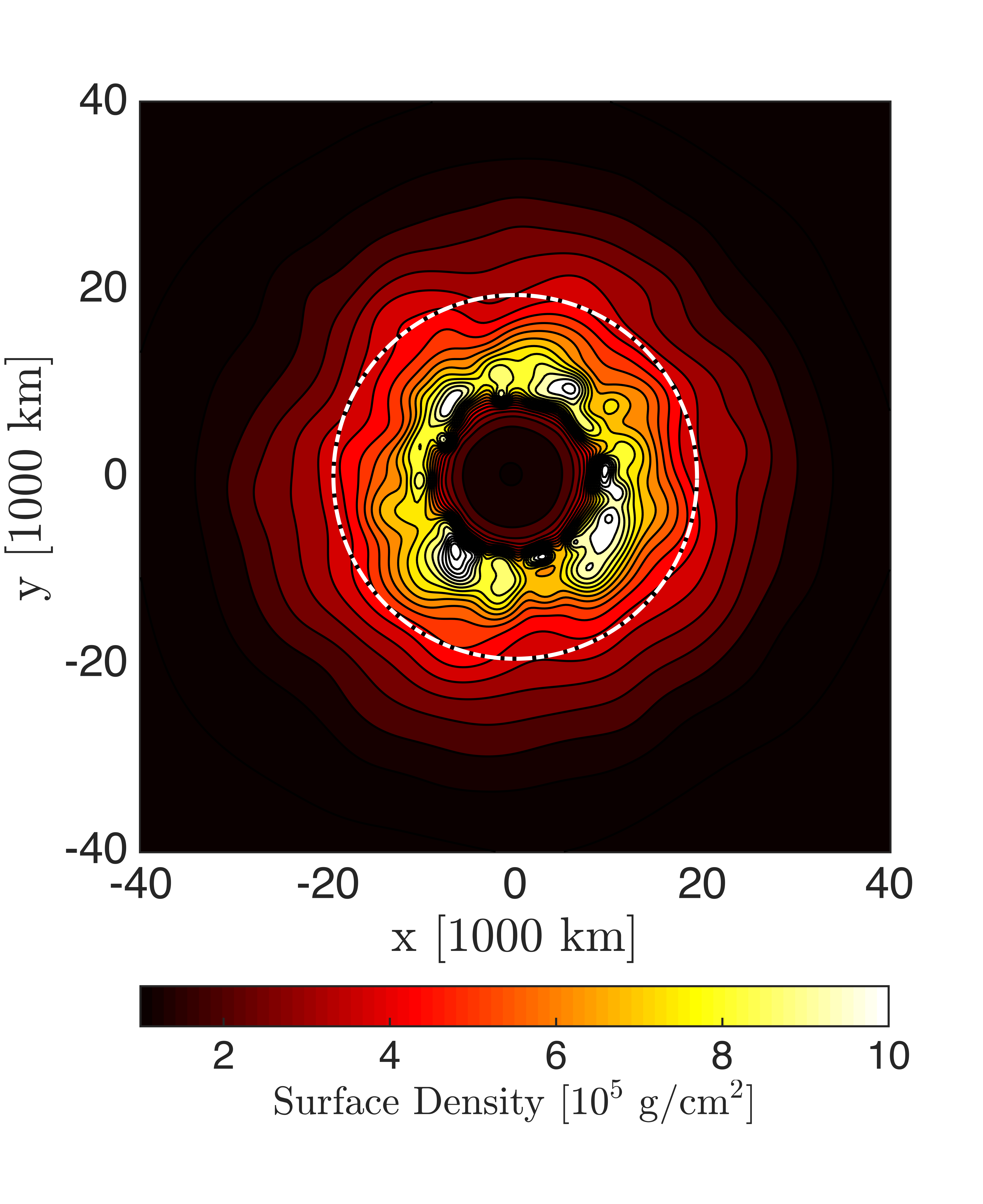}\includegraphics[clip,height=0.21\paperheight]{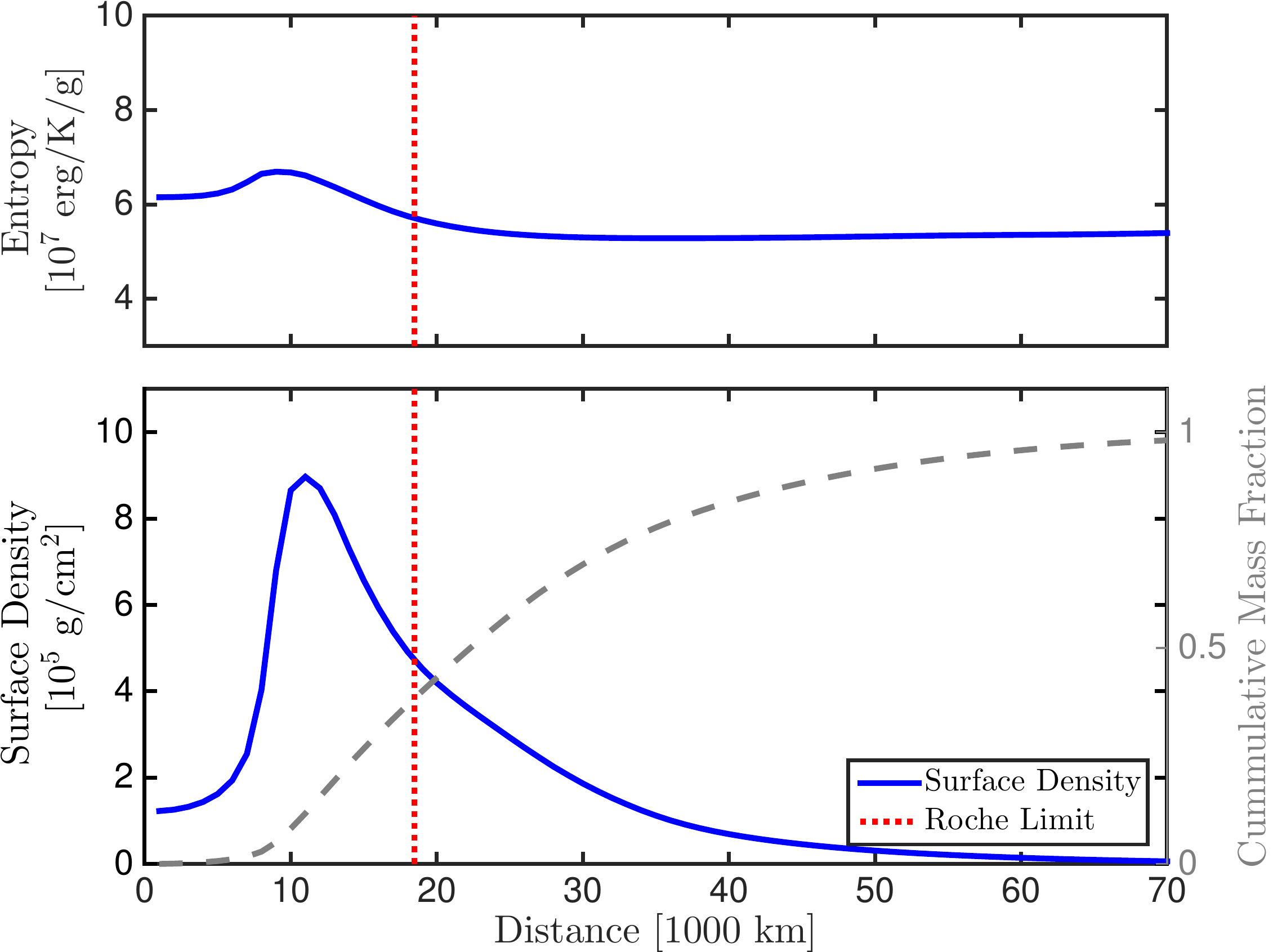}

\textbf{b)}\hspace{0.29\textwidth}\textbf{d)}\hspace{0.23\textwidth}

\includegraphics[scale=0.28]{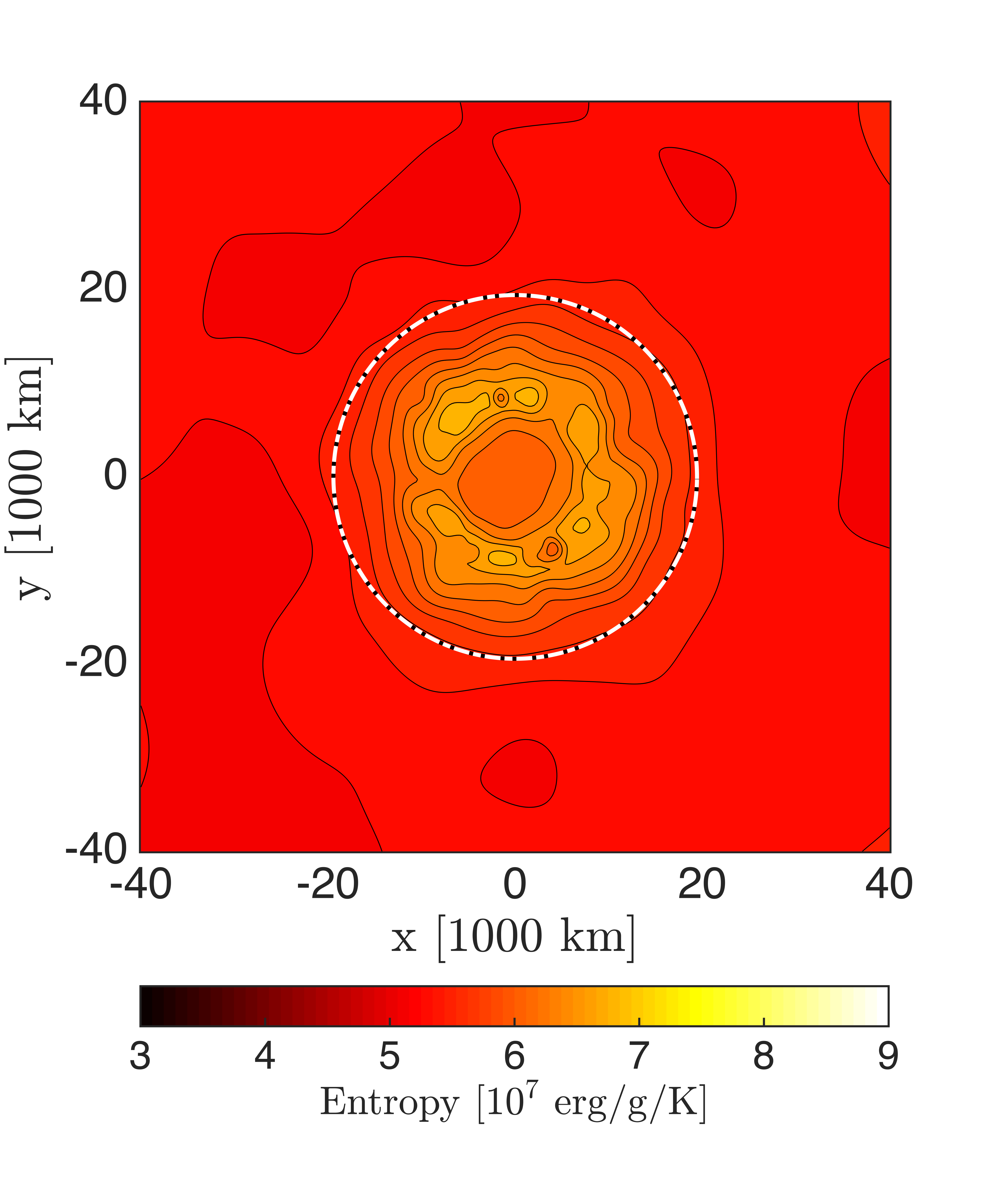}\hspace{0.014\textwidth}\includegraphics[scale=0.4]{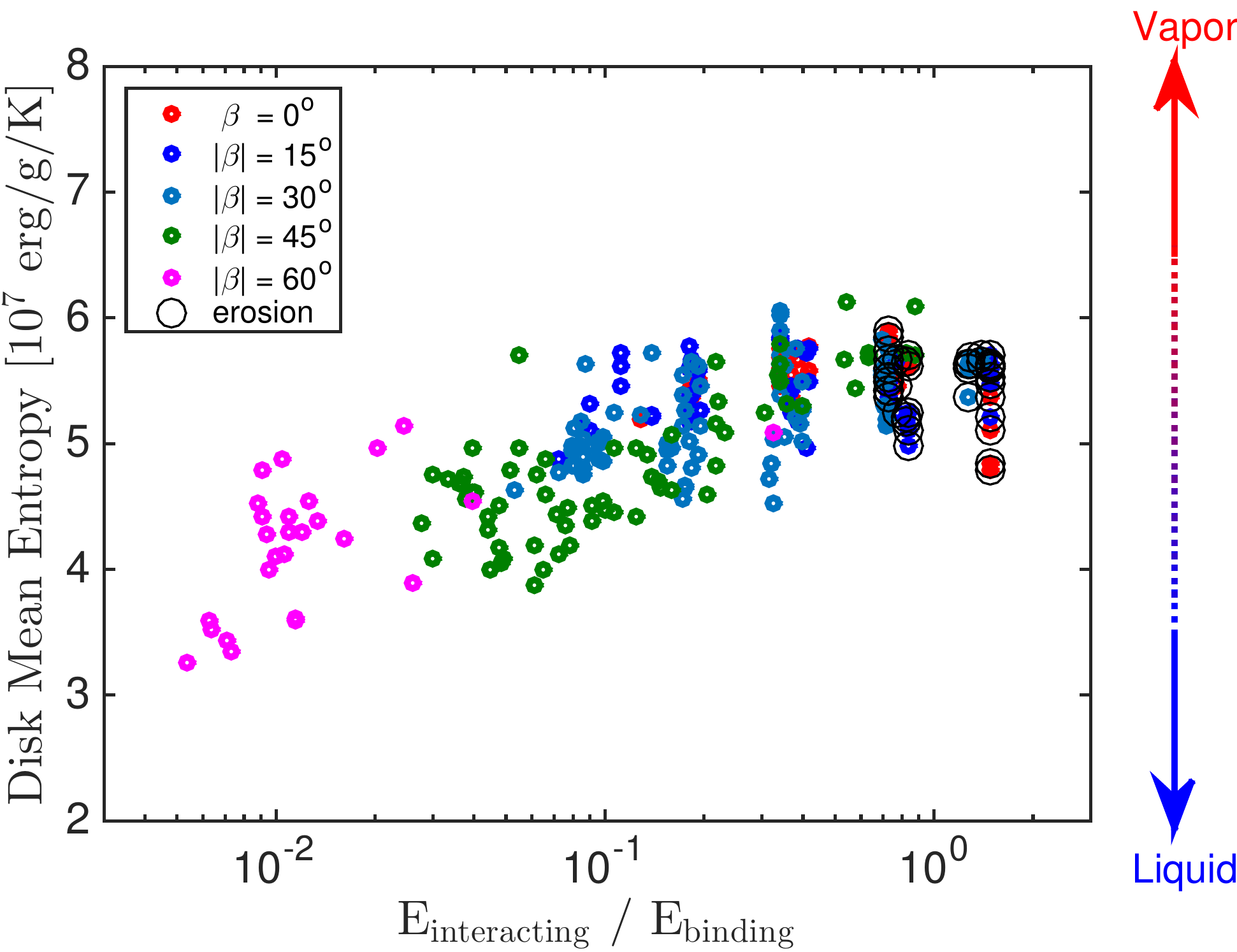}

\caption{\textbf{Disk structure post impact.} 2D structure of the a) surface
density b) entropy of the disk; c) Zonal average of surface density
and entropy after circularization of the disk vs. the distance from
the planet. The dashed lines in a-c) represent the Roche limit. The
initial conditions are the same as in Fig.~2. d) For our complete
set of simulations, mean entropy depends upon the energy of the interacting
mass (see Methods), normalized by the gravitational biding energy
of the target ($E_{{\rm binding}}=GM_{{\rm target}}^{2}/R_{{\rm target}}$).
Cases exhibiting planetary erosion (defined as $M_{{\rm planet}}\leq0.85M_{{\rm target}}$)
are indicated by black circles. }
\end{figure}

\section*{Implications for Moon Formation}

By releasing some of the constraints on the impact-generated debris
disk mass and system angular momentum due to individual impacts, the
multiple-impact scenario allows a significantly expanded range of
parameter space compared to previous Moon formation scenarios\cite{cuk2012making,canup2004simulations,Reufer2012296,Canup23112012}.
Freedom in impact geometry and velocity allows mining more material
from Earth, and the sum of such impact-generated moonlets may naturally
lead to the current values of the Earth-Moon system. Most terrestrial
rocks have similar ratios of $^{182}W/^{184}W$, however excess in
these ratios was found for Kostomuksha komatiites rocks dated at 2.8
billion year ago\cite{Touboul02032012}, suggesting an unmixed mantle
reservoir. Evidence of an unmixed mantle was also observed in noble
gas samples \cite{Mukhopadhyay:2012aa}. Efficient mantle mixing is
predicted for the single impact high angular momentum scenarios\cite{Nakajima2015286}
erasing any primordial heterogeneity that predated Moon formation.
Multiple smaller impacts promote preservation of primordial heterogeneity
of Earth's mantle, and potentially, also contribute to that of the
Moon \cite{Robinson2016244,Jutzi:2011aa}. 

We find that debris disks resulting from medium to large size impactors
($0.01-0.1M_{\oplus}$) have sufficient angular momentum and mass
to accrete a sub-lunar-size moonlet. We performed 1000 Monte Carlo
simulations of sequences of $N=$10, 20 and 30 impacts each, in order
to estimate the ability of multiple impacts to produce a Moon-like
satellite. The impact parameters were drawn from distributions previously
found in terrestrial formation dynamical studies\cite{Raymond2009644}.
With perfect accretionary mergers, approximately half the simulations
result in a moon mass that grows to its present value after $\sim20$
impacts (see Supplementary Material). We verified that the standard
deviation of the $\Delta^{17}{\rm O}$ difference between the growing
moon and target is reduced initially as $\sqrt{N}$, as expected.
After $\sim20$ impacts (roughly equivalent to the target mass) the
compositional difference stabilizes, with each growing body maintaining
a signature, owing to different proportions contributed by the incoming
sources. In the canonical giant impact scenario, only $\sim1-2\%$
of impactors lead to the observed compositional similarity (assuming
the same impactor composition distribution as described, and that
the impactor constitutes $60-80\%$ of the disk\cite{Canup2008518,Canup:2004aa}),
while in multiple impact scenarios the fractions of successful simulations
increases to 10\textquoteright s \%, with the precise values depending
on the assumed distribution. The range of conditions producing a Moon-mass
satellite with terrestrial composition is confined to narrow set of
conditions of fast ($V/V_{{\rm esc}}$$\sim4$), near head-on events
($\beta\sim0$). If multiple impacts are considered, additional components
with lower impact velocity are allowed, such as $V/V_{{\rm esc}}$$\sim2$
with $|\beta|<30^{o}$. This occurs to some extent because of the
reduction of the variance composition in averaging multiple components,
but also importantly, because there are regions in phase space where
the predicted moonlet masses are small and the compositions are Earth-like.
Such sub-lunar components can still provide the building blocks for
the final Moon, while in the single impact hypothesis they are rejected
due to their small mass.

Preimpact planetary rotation promotes moonlet formation by facilitating
the formation of a more massive disk from more loosely bound planetary
material, and by increasing the angular momentum of the disk, stabilizing
it against collapse. But the preimpact rotation rate is limited. Our
hydrodynamic simulations show that multiple impactors with isotropic
directions cannot accelerate the planet to breakup rotation due to
angular momentum drain by escaping material. The rotation of an Earth-like
target begins to saturate at a period $\sim5.9$ hours, in contrast
to previous calculations\cite{Agnor1999219} which assume perfect
merger. Faster rotation rates may be conceivably achieved by rare
single large events, but they are not expected to develop from the
sum of multiple collisions. This is validated by the Monte Carlo simulations
of impact sequences showing the vast majority of cases remain within
the range $-0.5\,\omega_{{\rm max}}\leq\omega\leq0.5\,\omega_{{\rm max}}$
(Supplementary Fig. 4-c). For sequences of 10, 20, and 30 impacts,
the planetary spin migrated outside this range for only 1, 7, and
13\% of the cases, respectively. Most cases have a final rotation
rate $-0.2\,\omega_{{\rm max}}\leq\omega\leq0.2\,\omega_{{\rm max}}$.
This neglects the change in planetary rotation due to moonlet tidal
evolution, that would generally decrease the spin rate.

The preferred scenario of the single-impact high angular momentum
case\cite{cuk2012making} invokes a small retrograde impactor that
produces a massive satellite due to the rapid initial rotation assumed
for the planet (near the breakup rate). Our conclusions, on the limited
planetary rotation gain and the resulting debris disk masses, point
to typical individual moonlet masses being considerably below the
present Moon mass. Specific scenarios that exceed one lunar mass may
be constructed. However, the broad range we find within the typical
parameter space that leads to sub-lunar mass disks, supports the notion
that the Earth's Moon was formed by the merger of multiple moonlets.

Satellite pairs formed by the same impact were found to be mostly
unstable\cite{Canup1999}, leading to moonlet-Earth collisions, moonlet-moonlet
mergers or scattering. High percentage of moonlet merger was found
for cases in which the inner moonlet is larger, hence tidally evolves
faster than the outer moonlet. The multiple impact scenario operates
on longer timescales, allowing even a smaller inner moonlet to reach
the Hill Radius of the outer moonlet in suitable time ($\sim100~{\rm Myr}$
for a 0.1 $M_{{\rm moon}}$ inner moonlet and a 0.9 $M_{{\rm moon}}$
outer moonlet at 30 $R_{{\rm Earth}}$, assuming current tidal values).
Moreover, the survival of an accreted moonlet depends also on the
planetary environment, as collisionless encounters of leftover planetesimals can
excite the satellite\textquoteright s eccentricity and possible loss\cite{Pahlevan:2015aa}.
Future work will address the new dynamics and merger efficiency of
the accreted moonlets to a final Moon. 

\section*{Methods}

We probe a broad phase space by simulating impact scenarios with ranging
values of the impactor\textquoteright s velocity, mass, angle, and
initial rotation of the target. We use Smoothed Particle Hydrodynamics
(SPH) to simulate impacts in the gravity dominated regime, using the
astrophysical code Gadget2\cite{springel2005cosmological}. In this
implementation, the individual evolution of spherically symmetric
particles is tracked in 3 dimensions. The spatial distribution of
each particle is defined by a spline density weighting function, known
as the kernel, and a characteristic radius, known as the smoothing
length. The functional form of the kernel does not change during the
simulation, but the smoothing length of each particle is varied in
order to maintain a constant desired number of overlapping particles,
or neighbors. This procedure allows low-density regions to be resolved,
although spatial resolution is reduced. The kinematic state of each
particle is evolved due to gravity, compressional heating/expansional
cooling, and shock dissipation. Material strength and fracture are
neglected, and for the simulated time of the impact ($\sim$a day)
radiative processes may be ignored. A standard prescription of artificial
viscosity is used in order to mimic shock dissipation. 

We employ a tabulated equation of state which was generated using
the well-tested semi-analytical equation of state M-ANEOS\cite{melosh2007hydrocode}
with forsterite comprising the mantle and iron comprising the core.

We performed over 800 simulations consisting of $10^{5}$ SPH particles
to span the parameter space and a smaller number of high resolution
simulations with $10^{6}$ particles to better define the characteristics
of the disk.

For ease of comparison with previous work\cite{cuk2012making,Canup2008518},
all impacts are assumed to occur in the equatorial plane of the target.
The runs cover 24 hours of simulation time after impact.\foreignlanguage{american}{\textit{
}In order to test consistency of our hydrodynamic code and analysis
algorithms, we reproduced previously published scenarios\cite{cuk2012making}
and obtained small differences (\textit{e.g.}, $<5\%$ in disk mass). }

We use a target with a mass of $\sim1$ Earth mass because we\foreignlanguage{american}{ assume
that the Moon-forming impacts occur near the final stages of the planet's
accretion} when the target mass changes by a relatively small amount
with each impact. In off-axis impacts, the fraction of the impactor
mass whose projected initial trajectory intersects the planet is referred
to as the interacting mass, $M_{{\rm interacting}}$, with kinetic
energy $E_{{\rm interacting}}$.

We validated our assumption that previous accreted moonlets do not
change the initial outcome of the disk, by simulating impacts on targets
with an orbiting moon. We chose a Mars-sized orbiting body with a
range of semimajor axes, from the closest stable orbit in the SPH
simulation to the current location of the Moon ($(1.2-20)R_{{\rm Roche}})$
and an equal sized impactor. The results show that a pre-existing
moonlet does not affect the disk (mass and composition) as long as
it is not close to the Roche limit ($>3\,R_{{\rm Roche}})$. For very
close-in satellites, the impact can induce collision in the planet-moon
system. For very distant satellites, the impact can excite the moons
into escaping trajectories. Massive moons, like the one tested, will
migrate quickly after their formation due to tidal dissipation ($\sim O(3)\ {\rm yr}$
from $R_{{\rm Roche}}$ to $3\,R_{{\rm Roche}}$ for current dissipation
parameters), hence preexisting moons will be distant enough before
the next impactor so that we may restrict our attention to simulating
only two bodies. We note that, we test only the first $\sim$24 hours
after the impact, while the longer disk evolution can be affected
by resonances and angular momentum transfer with pre-existing moons\cite{Charnoz:2010aa}. 

\subsection*{\label{subsec:Post-impact-analysis}Impact Characteristics }

We characterize the impact event by determining the mass of the disk
produced, mass of the planet and angular momentum and composition
of the components. After each simulation we follow an iterative procedure\cite{Canup23112012,canup2001scaling}
to classify particles into one of three categories: planet particles
whose angular momentum is insufficient to escape the gravitational
pull of the planet; disk particles whose orbital pericenter is outside
the planetary radius, and escaping particles, which are gravitationally
unbound. We chose to classify disk particles as those with periapsis
distance larger than the equatorial radius of the planet, as opposed
to using the semimajor axis equivalent distance ($a_{{\rm eq}}=l_{z}^{2}/GM_{{\rm planet}}$),
because the latter overestimates the disk mass in cases where the
disk particles are not vaporized, eccentric, and not likely to further
gravitationally interact. (We note that in high energy and high angular
momentum cases\cite{LockStewart}, this distinction is irrelevant).
In this classification we neglect pressure gradients as particles
are assumed to travel on Keplerian orbit, as would solid particles. 

The compositional difference between the silicates in the final planet
and disk is defined as\cite{Canup23112012,cuk2012making}:

\begin{equation}
\delta f_{{\rm T}}=\left[f_{{\rm disk,tar}}/f_{{\rm planet,tar}}-1\right]\times100,\label{eq:DeltaFrac}
\end{equation}
 where $f_{{\rm disk,tar}}$ and $f_{{\rm planet,tar}}$ are the mass
fractions of silicate originating from the target in the disk and
the final planet, respectively. A value of $\delta f_{{\rm T}}<0$
implies that the disk contains more material derived from the impactor
than from the target. If the impactors had the same isotopic composition
as Mars then the compositional constraint allows only $|\delta f_{{\rm T}}|<2$\cite{Pahlevan:2007aa},
however the projectile may have been more similar to Earth than to
Mars \cite{Mastrobuono-Battisti:2015aa}. The sum of multiple disks
will contribute to the final moon's composition, so each disk can
vary to some extent while maintaining an average of small $|\delta f_{{\rm T}}|$. 

The slower accretion of the moonlet from the disk is not simulated
but estimated from the calculated disk mass, $M_{{\rm disk}}$, and
its angular momentum, $L_{{\rm disk}}$, using results from previous
N-Body simulations \cite{ida1997lunar}:

\begin{equation}
M_{{\rm sat}}=1.9\cdot\frac{L_{{\rm disk}}}{\sqrt{GM_{{\rm planet}}a_{{\rm R}}}}-1.15M_{{\rm disk}}-1.9M_{{\rm {\rm esc}}},\label{eq:M_sat}
\end{equation}

where $G$ is the gravitational constant. This relation is an estimate
for the satellite mass as the exact accretion efficiency is still
uncertain. We assume that the escaped mass during the accretion process
is negligible ($M_{{\rm esc}}=0$) as in previous studies\cite{Canup23112012,cuk2012making}.
The equation is not valid for disks with high specific angular momentum
(yielding $M_{{\rm sat}}>M_{{\rm disk}}$) that correspond to a moon
accretion at distances greater than $1.3\,R_{{\rm Roche}}$. We neglect
the disks that yield a negative angular momentum compared with Earth,
as a moonlet in a retrograde orbit will experience tidal acceleration
that will lower the semimajor axis until it collides with the planet. 

We assume that initially eccentric and inclined disks are circularized
and flattened as collisions between particles in the disk will lead
in a few orbital periods to decrease in orbital eccentricities and
inclinations while angular momentum is conserved. The final semimajor
axis of each disk particle is $a_{{\rm eq}}=l_{z}^{2}/GM_{{\rm planet}}$,
where $l_{z}$ is the specific angular momentum of the particle normal
to the equatorial plane of the planet. By lowering all the disk particles
to the equatorial plane, assigning them a random phase and smoothing
them according to the kernel function, we calculate the surface density
and specific entropy of the disk at any given point. 

\subsection*{Monte Carlo Simulations}

We performed Monte Carlo simulations of impact sequences by drawing
impacts parameters \foreignlanguage{american}{(mass ratio, $V_{{\rm imp}}/V_{{\rm esc}}$,
impact angle \textgreek{b})} within our phase space of interest and
distributions based on previous terrestrial formation studies \cite{Raymond2009644}.
The angle distribution is symmetric, allowing retrograde and prograde
impacts. Due to the uncertainty in the impact parameters, we also
separately examined sequences of high velocities ($2.5-4V_{{\rm esc}}$)
and low velocities ($1-2.5V_{{\rm esc}}$) impactors. \foreignlanguage{american}{The
impactor mass ratio of interest in this work is 1-9\%, the upper limit
corresponding to the approximate isolation mass\cite{Kokubo2006}
(the protoplanetary mass expected to accrete by the oligarchic growth
in the early planetesimal disk) of the inner solar system. Although
rare larger events are possible, our goal here is to demonstrate the
feasibility of a scenario consisting of multiple smaller events.}
Each impactor has a distinct compositional signature drawn from a
normal distribution with a standard deviation of a Mars-like composition\cite{Pahlevan:2007aa}. 

For every simulated impact, we perform a linear interpolation within
our phase space to determine post impact conditions such as planetary
rotation spin, moonlet mass, iron disk content and the impactor contribution
to the disk and planet. The planetary spin and signature is evolved
from one impact to the next. The final Moon's characteristics are
calculated assuming perfect moonlet accretion and the difference between
the planet's silicate composition and that of the growing moon's is
tested. By assuming that the the iron fraction in the disk is preserved
in the accreted moonlet, we estimate the iron fraction in the final
moon. The final angular momentum of the system is obtained from the
sum of the individual disk angular momenta with Earth's final value.
Perfect accretion of moonlets is assumed as a simplification, as the
details and probabilities of various multi moonlet evolutionary paths
are yet unknown.

\section*{Code availability}

The code used to generate the hydrodynamic simulations can be accessed
\textit{https://wwwmpa.mpa-garching.mpg.de/gadget/}. 

\section*{Data availability }

The data that support the findings of this study are available from
the corresponding author upon request. 

\bibliographystyle{naturemag}

\section*{Acknowledgments}

We thank S. Stewart and R. Citron for providing guidance on the computational
code, as well as A. Mastrobuono-Battisti for providing the data used
for the Monte Carlo simulations. This project was supported by the
Minerva Center for Life Under Extreme Planetary Conditions as well
as by the I-CORE Program of the PBC and ISF (Center No. 1829/12).
RR is grateful to the Israel Ministry of Science, Technology and Space
for their Shulamit Aloni fellowship. HBP also acknowledges support
from the Israel-US bi-national science foundation, BSF grant number
2012384, and the European union career integration grant \textquotedblleft GRAND\textquotedblright .

\section*{Author Contribution}

RR performed the SPH simulations and their analysis with guidance
by OA. HBP suggested the multiple impact idea. All authors contributed
to discussions, interpretations and writing.

\section*{Competing Financial Interest}

The authors declare no competing financial interests.

\section*{}

\textbf{ }

\end{document}


\title{Supplementary material :\textbf{ }\foreignlanguage{english}{A Multiple
Impact Origin for the Moon}}

\author{Raluca Rufu$^{1*}$, Oded Aharonson$^{1}$,\foreignlanguage{english}{
Hagai B. Perets$^{2}$}}
\maketitle

\lyxaddress{$\!$$^{1}$Weizmann Institute of Science, Department of Earth and
Planetary Sciences, Center for Planetary Science, Rehovot 76100, Israel}
\selectlanguage{english}%

\lyxaddress{$^{2}$ Technion Israel Institute of Technology, Physics Department,
Haifa 32000, Israel.}
\selectlanguage{american}%
\begin{center}
{*} Contact Information: e-mail: Raluca.Rufu@weizmann.ac.il
\par\end{center}

\tableofcontents{}

\hfill{}

Other Supplementary Material for this manuscript includes the following:

Movie S1

Table of Results

\newpage{}

\selectlanguage{english}%
\selectlanguage{american}%

\section{Supplemental Methods}

\subsection{Equation of State}

We employ a tabulated equation of state which was generated using
the well-tested semi-analytical equation of state M-ANEOS \cite{melosh2007hydrocode}
with forsterite comprising the mantle and iron comprising the core.
ANEOS uses the temperature and density as independent variables and
thermodynamic quantities are derived from the Helmontz free energy,
described by the sum of the zero-temperature free energy, thermal
energy of the collective motion of the nuclei, and an electronic ionization
term. The equation of state describes a mixed vapor-fluid state in
a single SPH particle, assuming that the phases are in temperature
and pressure equilibrium, therefore the mass fraction of each state
can be calculated. M-ANEOS is a revised version of ANEOS \cite{thompson1990aneos}
which adds another 8 parameters (a total of 45 parameters) to describe
diatomic molecular vapor that previously treated all vapor as monoatomic
species instead of molecular gas phases (mostly ${\rm SiO_{2}}$ and
${\rm O_{2}}$), hence the entropy required for vaporization was overestimated.
It should be noted that M-ANEOS (and ANEOS) cannot account for metastable
states as perfect thermodynamic equilibrium and perfect reversibility
is assumed.

\subsection{Impact simulations}

We performed over 500 simulations consisting of $N\sim10^{5}$ SPH
particles to span the parameter space and a smaller number of high
resolution simulations with $N\sim10^{6}$ particles to better define
the characteristics of the disk. We choose parameters for the impactor
mass ratio $\gamma$, speed $V_{{\rm imp}}$, angle $\beta$, and
planetary rotation $\omega$ (Supplementary Table \ref{tab: Initial Condition Table}).
\begin{table}[p]
\begin{centering}
\begin{tabular}{l|l|l}
\noalign{\vskip0.1cm}
 & \textbf{Parameter} & \textbf{Values}\tabularnewline[0.1cm]
\hline 
\noalign{\vskip0.1cm}
mass ratio & \textbf{$\mathbf{\gamma}$} & 0.010, 0.024, 0.053, 0.091\tabularnewline[0.1cm]
\hline 
\noalign{\vskip0.1cm}
impact velocity & \textbf{\large{}$V_{{\rm imp}}/V_{{\rm esc}}$} & 1.0, 1.1, 1.4, 2.0, 3.0, 4.0\tabularnewline[0.1cm]
\hline 
\noalign{\vskip0.1cm}
impact angle & {\large{}$\mathbf{\beta\:[\:^{\circ}\:]}$} & 0, $\pm$15, $\pm$30, $\pm$45, $\pm$60\tabularnewline[0.1cm]
\hline 
\noalign{\vskip0.1cm}
 planetary rotation & \textbf{\large{}$\omega/\omega_{{\rm max}}$} & 0.00, 0.10, 0.25, 0.50\tabularnewline[0.1cm]
\hline 
\end{tabular}
\par\end{centering}
 \renewcommand\tablename {Supplementary Table} 

\caption{\foreignlanguage{english}{\label{tab: Initial Condition Table}\textbf{Values of the different
initial conditions assumed}. The parameters are represented by $\gamma$,
the mass ratio between the impactor and total mass, $V_{{\rm imp}}$,
the impact velocity normalized by the mutual escape velocity at contact,
$\beta$, the impact angle defined as the angle between the centers
of the bodies at contact and the velocity of the impactor, $\omega$
initial rotation rate of the proto-Earth normalized by $\omega_{{\rm max}}$
, the maximum rotation rate before break-up. }}
\end{table}

The colliding bodies contain 30\% core and 70\% mantle material, similar
to current Earth, and are generated with isentropic thermal profiles
with a surface temperature of 2000K comparable to a 'warm start' in
previous studies \cite{Canup23112012}. The objects are simulated
in isolation for 24 hours to allow their initial relaxation and to
establish gravitational equilibrium. Rotation is added to a stable
planet and is simulated again in order to re-establish equilibrium.

We characterize the impact event by determining the mass of the resulting
disk and planet, as well as the angular momentum and composition of
the components. After each simulation we follow an interactive procedure
\cite{canup2001scaling,canup2004simulations} to classify particles
into one of three categories: planet particles whose angular momentum
is insufficient to escape the gravitational pull of the planet; disk
particles, distinguished from the planet, by their orbital periapses
falling outside the planetary radius, and escaped particles, which
are gravitationally unbound. In this classification we neglect pressure
gradients as particles are assumed to travel on Keplerian orbit, as
would be the case for non-interacting solid particles. An initial
guess is made for the planet's mass, $M_{{\rm planet}}$, and flatness,
$f$. Assuming a terrestrial density of $\rho_{\oplus}=5.5\:{\rm g/cm^{3}}$,
the oblate spheroid equatorial radius is:

\begin{equation}
R_{{\rm eq}}=\left(\frac{3}{4\pi\rho_{\oplus}}M_{{\rm planet}}\frac{1}{1-f}\right)^{1/3}.
\end{equation}
An alternative to the condition comparing $R_{{\rm eq}}$ to the periapses
distance is to compare it to the semimajor-axis equivalent distance
($a_{{\rm eq}}=l_{z}^{2}/GM_{{\rm planet}}$). Although for low eccentricity
trajectories the two conditions were found to agree \cite{canup2004simulations},
we found that in some of the conditions they may differ by as much
as a factor of three in the total disk mass. This difference plays
a more significant role in less massive disks, such as the ones considered
here, as the energy dissipation time scale is larger than the collision
time of the particle with the planet. We therefore choose the periapses
condition as it is the more rigorous threshold in our simulations.
To treat particle aggregates, we search for particles outside the
planetary radius with $\rho>1{\rm \,g/cm^{3}}$, and sort them into
density 'clumps'. Each group is assigned a uniform classification
according to the dynamics of its center of mass, similar to past treatments
\cite{Nakajima2014259}.

The mass and angular momentum of the ejected particles (disk+escaped)
are used to calculate a new estimate for the planet's mass and its
rotation ($\omega_{{\rm planet}}$). A new flatness parameter is calculated
using:

\begin{equation}
f=\frac{5}{2}\left(\frac{\omega_{{\rm planet}}}{\omega_{{\rm max}}}\right)^{2}/\text{\ensuremath{\left[1+\left(\frac{5}{2}-\frac{15K}{4}\right)^{2}\right]}}\mbox{,}\label{newflatness}
\end{equation}

where $\omega_{{\rm max}}$ is the maximum rotation before break-up
and $K$ is the gyration constant (for Earth $K=0.335$, for a uniform
density sphere $K=\frac{2}{5}$ ). This value of $f$ and $M_{{\rm planet}}$
is then used to calculate a new $R_{{\rm eq}}$. The algorithm is
repeated until convergence.

\section{Supplemental Results}

\subsection{Debris Disk}

\begin{figure}[p]
\begin{centering}
\includegraphics[scale=0.4]{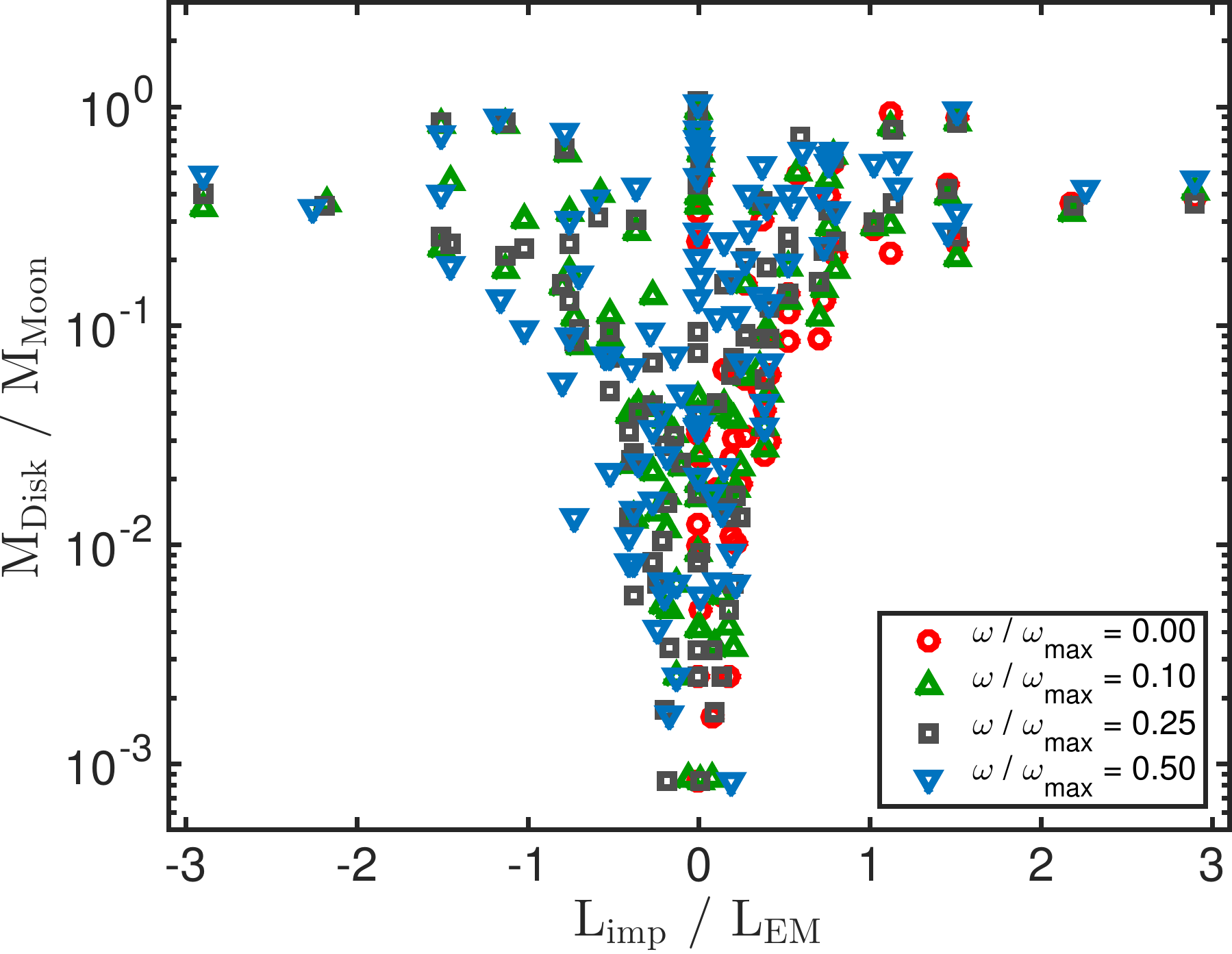}
\par\end{centering}
 \renewcommand\figurename {Supplementary Figure} 
\centering{}\caption{\foreignlanguage{english}{\label{fig:MassDiskVsL}\textbf{Disk mass as a function of impact
angular momentum} \textbf{for low impact angles ($|\beta|\leq30^{o}$)}.
The different markers style represents different initial rotations. }}
\end{figure}

The general trend observed in the disk mass is that the non-head-on
($\beta\geq15^{o}$) and prograde impacts ($L_{{\rm imp}}>0$, where
the impactor's angular momentum is in the same direction as that due
to the planetary spin) result in disks increasing in mass as a function
of angular momentum, until a maximum is reached and excess angular
momentum causes material to escape the system at the expense of the
disk (Supplementary Figure \ref{fig:MassDiskVsL}). The maximum disk
mass is higher with increasing rotation rate, as material is more
loosely bound to the planet and is ejected from the surface more easily.
For the retrograde impacts ($L_{{\rm imp}}<0$), disk mass still increases
with absolute angular momentum, however peaks at a reduced value.\foreignlanguage{english}{
The mass of the disk does increase for a prograde impact by a factor
of 2 relative to retrograde (keeping other parameters constant), because
the latter has a lower total angular momentum, and the disk is correspondingly
less massive (}Supplementary\foreignlanguage{english}{ Figure 1) .}

The energetics of the disk will determine the initial thermal state
of each moonlet. Because the time between impacts is short, of order
Myrs, moonlets may retain much of this initial heat (and associated
melt) until subsequent merger.

The silicon vapor fraction is calculated by summing the masses of
the particles in a single vapor phase and the particles in the two
phase state multiplied by the vapor fraction for each particle ($X_{{\rm vap_{i}}}$)
using the relation: 
\begin{equation}
X_{{\rm vap_{i}}}=\frac{S_{i}-S_{{\rm liq}}(T_{i})}{S_{{\rm vap}}(T_{i})-S_{{\rm liq}}(T_{i})},
\end{equation}
where $S_{i}$ is the specific entropy of the SPH particle, $S_{{\rm liq}}(T_{i})$
and $S_{{\rm vap}}(T_{i})$ are the specific entropies of the liquid
and vapor phase boundary, respectively, at temperature $T_{i}$ of
the particle \textit{i}.

The silicon vapor fraction in the disk (shown in Supplementary Figure
\ref{fig:Disk-thermodynamics-as}) ranges from $\sim10-70\%$ with
low angle impacts angles exhibiting the maximum value.

\selectlanguage{english}%
Previous work \cite{Wada2006} of high-resolution Moon forming impacts
using grid-based simulations, suggests that a disk composed of mainly
silica vapor will not be sufficiently stable to accrete into a large
Moon owing to fast angular momentum transfer associated with spiral
shocks. However, the only tested scenario was similar to the canonical
impact \cite{Canup:2004aa} and the two polytrope equations of state
chosen are not able to mimic the phase changes in the mantle \cite{Nakajima2014259}.
Future work is still needed to determine whether high vapor fraction
disks are suitable for Moon forming scenarios, as well as to account
for lower disk mass and preexisting moons in current accretion models
\cite{Salmon20130256}.

\selectlanguage{american}%
\begin{figure}
\begin{centering}
\includegraphics[scale=0.4]{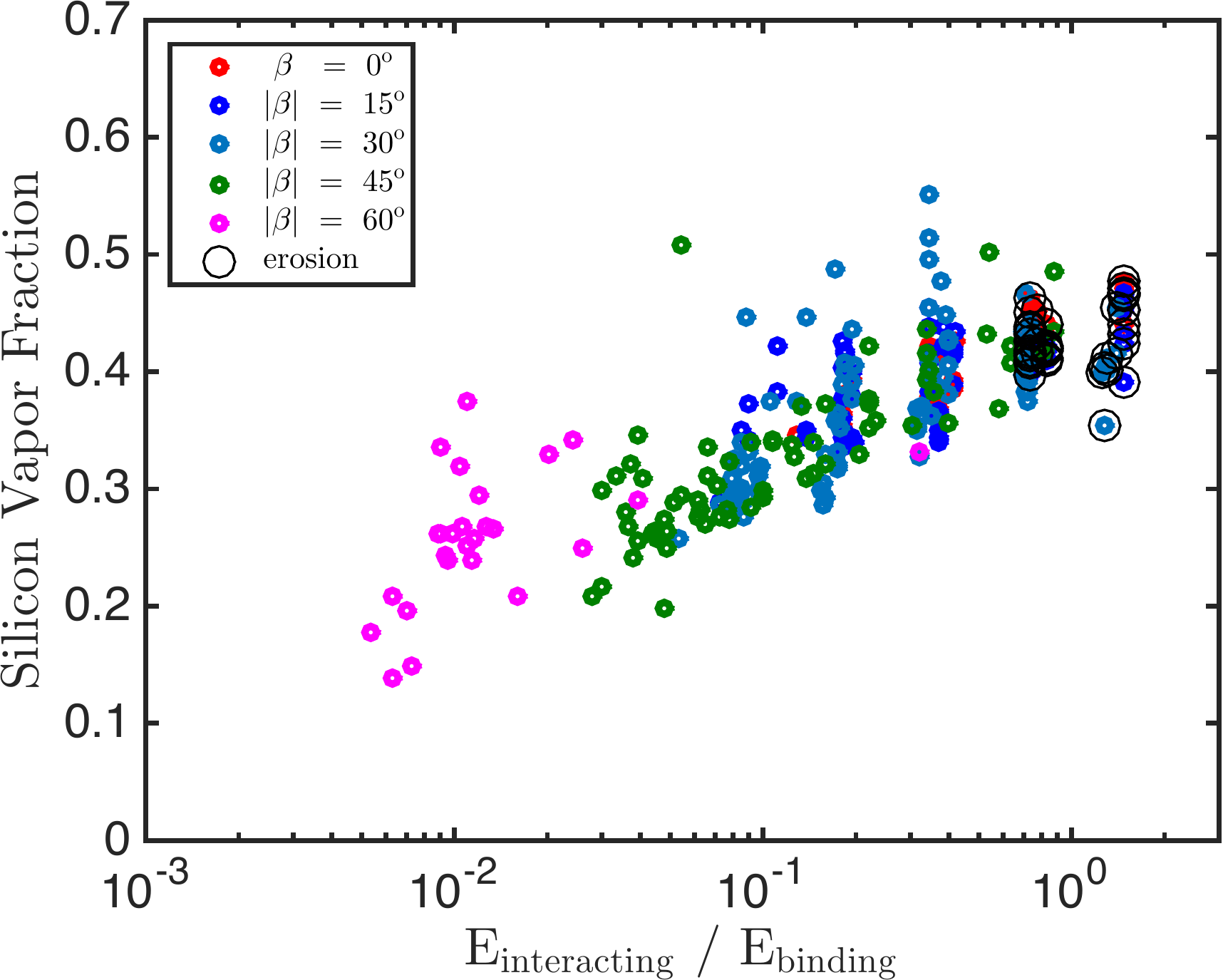}
\par\end{centering}
 \renewcommand\figurename {Supplementary Figure} 

\caption{\foreignlanguage{english}{\textbf{\label{fig:Disk-thermodynamics-as}Disk thermodynamics as
a function of impact energy.} Vapor fraction as a function of the
energy of the interacting mass (defined in the Methods section in
the main text), normalized by the gravitational biding energy of the
target ($E_{{\rm binding}}=GM_{{\rm target}}^{2}/R_{{\rm target}}$
). Cases exhibiting planetary erosion (defined as $M_{{\rm planet}}\leq0.85M_{{\rm target}}$)
are indicated by black circles. }}
\end{figure}

\subsection{Planetary Erosion }

\begin{figure}
\begin{centering}
\includegraphics[scale=0.4]{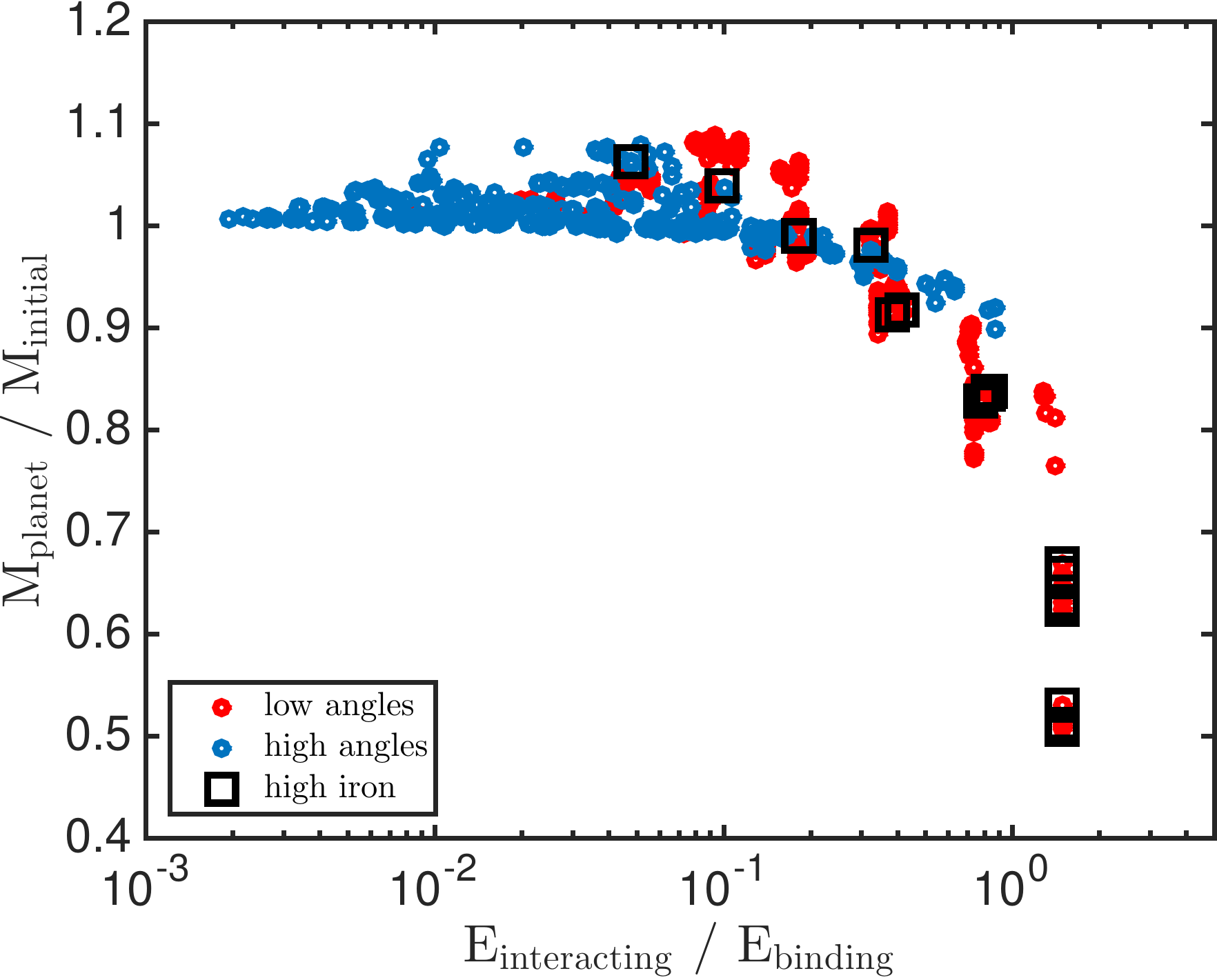}
\par\end{centering}
\begin{centering}
 \renewcommand\figurename {Supplementary Figure} 
\par\end{centering}
\centering{}\caption{\foreignlanguage{english}{\label{MPlanetVsEnergy}\textbf{Mass of the planet after impact as
a function of the energy of the interacting mass}. The interacting
mass is defined as the total energy of the impact scaled by the fraction
$M_{{\rm interacting}}/M_{{\rm impactor}}$ and is normalized by the
binding energy of the target ($GM_{{\rm target}}^{2}/R_{{\rm target}}$
). The blue circles represent impacts with high angles ($|\beta|>30^{o}$),
the red represent the low angles ($|\beta|\leq30^{o}$). The black
squares represent disk with a larger iron content than the Moon's
core. }}
\end{figure}

The impact parameter space is mostly composed of partial accretion
impacts. Extremely high energy impacts caused severe erosion to the
planet and contaminated the disk with core target material. In these
cases, the iron fraction of the disk is larger than the estimated
amount in the present Moon (Supplementary Figure \ref{MPlanetVsEnergy}).\textbf{
}Also, since mantle material is preferentially ejected relative to
the core, the iron fraction in the target grows (up to 60\%) for increasingly
erosional collisions. For these cases to be relevant for Moon formation
scenarios, we would need a smaller initial cores.

\subsection{Monte Carlo Simulations}

We perform monte Carlo simulations of impact sequences by drawing
impacts parameters from our phase space (mass ratio, $V_{{\rm imp}}/V_{{\rm esc}}$
, impact angle $\beta$) with similar distributions found in previous
terrestrial formation studies {[}\cite{Raymond2009644}. We also separately
examined sequences of high velocities impactors (\foreignlanguage{english}{$2.5-4\,V_{{\rm esc}}$})
and low velocities (\foreignlanguage{english}{$1-2.5\,V_{{\rm esc}}$})
due to uncertainty in the distribution of impact parameters. 

For a sequence of 20 impacts, the final Moon is massive enough ($\geq M_{{\rm moon}}$)
in 40\% of the cases. 17\% of simulations with sequences of 20 impacts
have a signature difference small enough to reproduce the oxygen isotope
signature difference in the Earth-Moon system ($\Delta^{17}{\rm O\leq0.005\permil}$).
For a population of high velocity impactors (\foreignlanguage{english}{$2.5-4\,V_{{\rm esc}}$}),
the compositional similarity increases to 36\%. This population of
impactors tends to excavate more planetary material, and additionally
in some cases a large part of the impactor escapes. In contrast, for
a population of low velocities impactors (\foreignlanguage{english}{$1-2.5\,V_{{\rm esc}}$}),
the compositional similarity is achieved in 13\% of the cases. This
population of impactors is often in the part of phase space where
graze and merge is common, contributing more impactor material to
more massive disks than in the faster regime. 

The iron fraction of the final Moon was less than 10\% (the estimated
upper limit \cite{canup2004simulations} for more than 75\% (98\%/60\%,
for high/low impactor velocity subsets) of the cases. The absolute
values of the final angular momentum of the systems have a median
of $0.5\,L$, and \textasciitilde{}10\% of cases (<1\%/17\% for high/low
velocity subsets) have a final value exceeding $L_{{\rm EM}}$. 

The final angular momentum resulting from multiple impacts is arbitrarily
set by the last few impactors. A value greater than $L_{{\rm EM}}$,
is reached throughout the evolution of a 20 impact sequence in \textasciitilde{}50\%
of the cases, making the $1\,L_{{\rm EM}}$ value feasible in this
scenario. Importantly, further changes to the total angular momentum
are possible by dynamical processes including the evection resonance
or an associated limit cycle \cite{Wisdom2015138}, moonlet escape
reducing the angular momentum, or semicollisional accretion \cite{schlichting2007effect}
increasing the angular momentum. 

\begin{figure}
\hspace{0.07\textwidth}a)\hspace{0.27\textwidth}b)\hspace{0.26\textwidth}c)
\begin{singlespace}
\begin{centering}
\includegraphics[clip,scale=0.25]{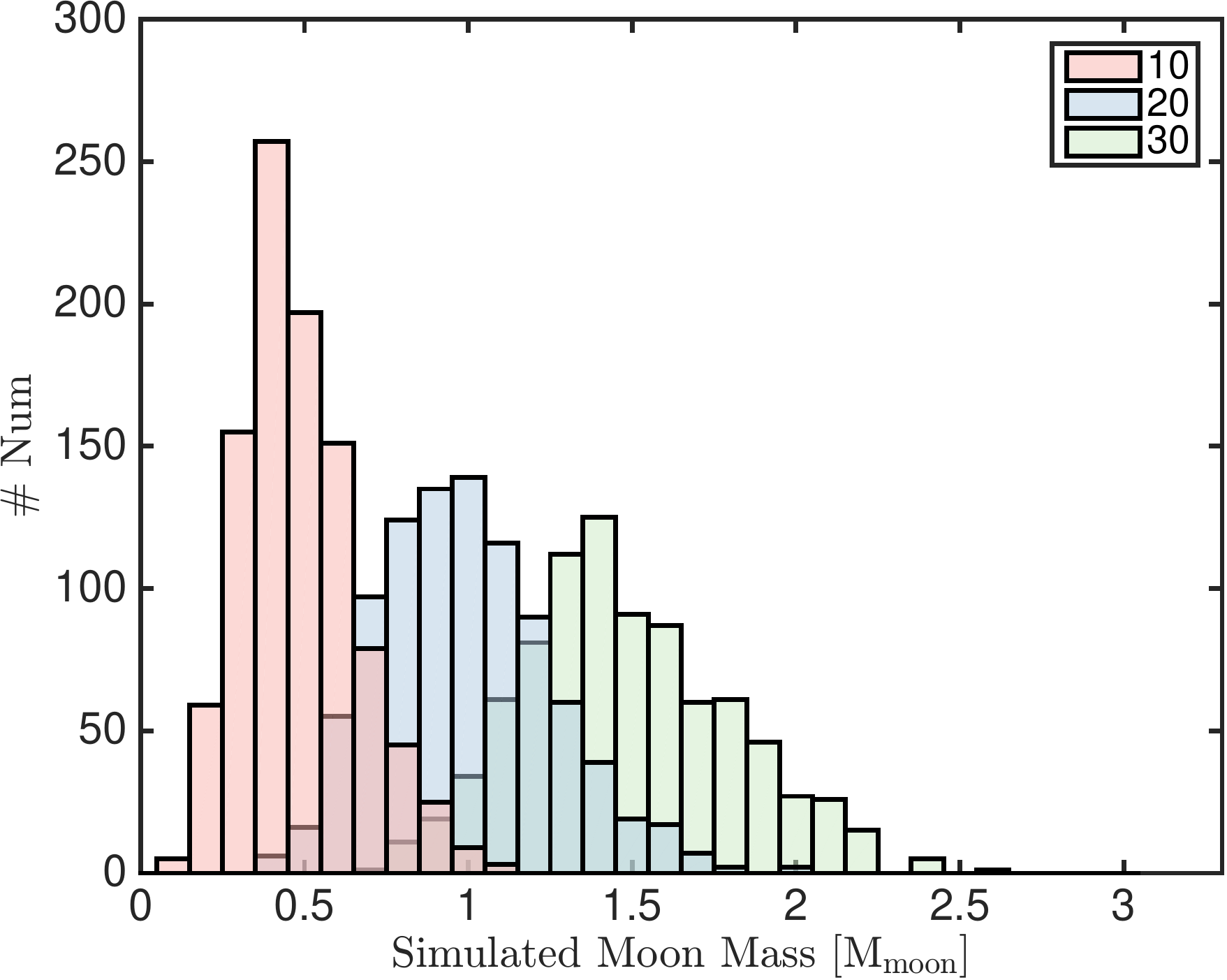}\includegraphics[clip,scale=0.25]{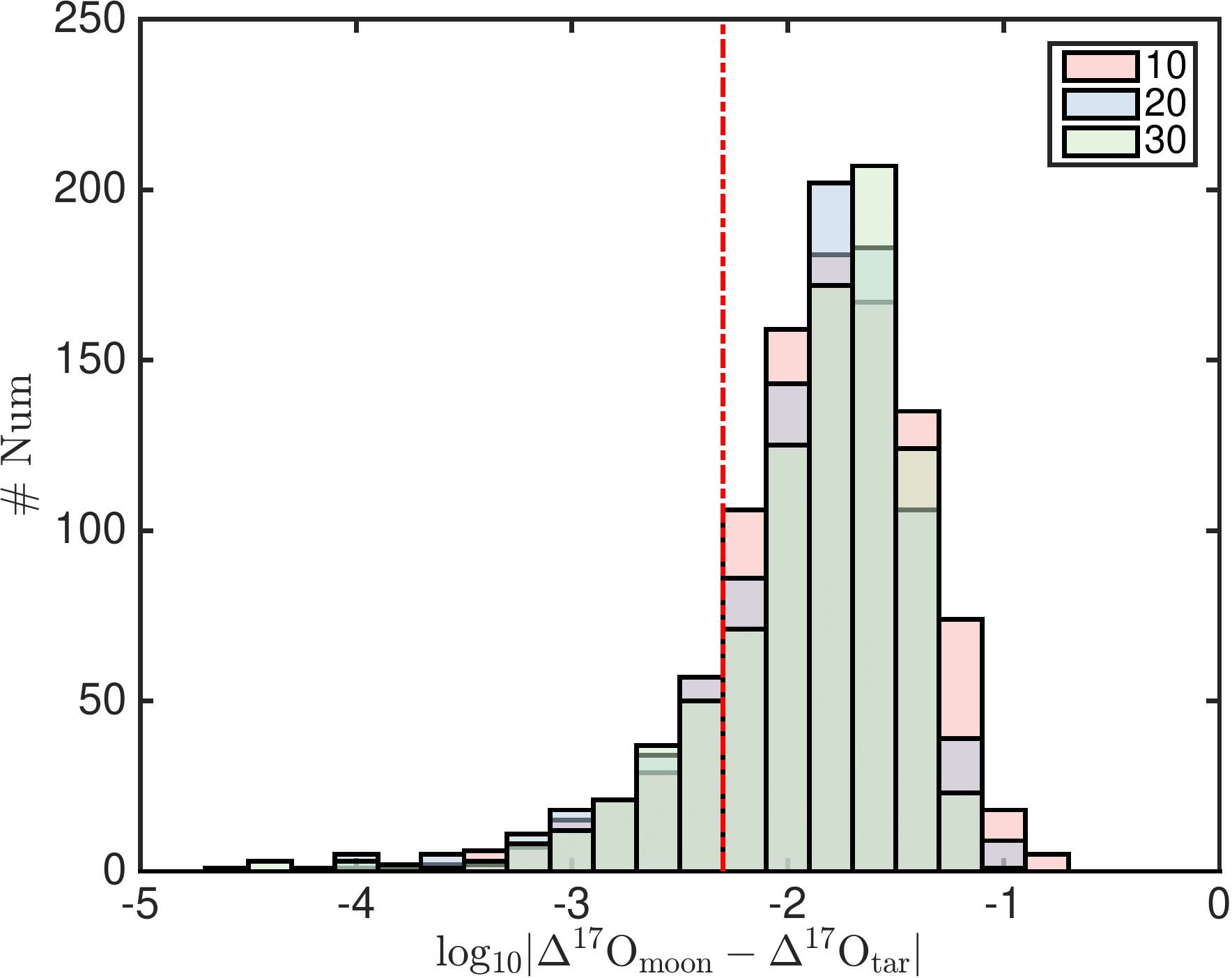}\includegraphics[clip,scale=0.25]{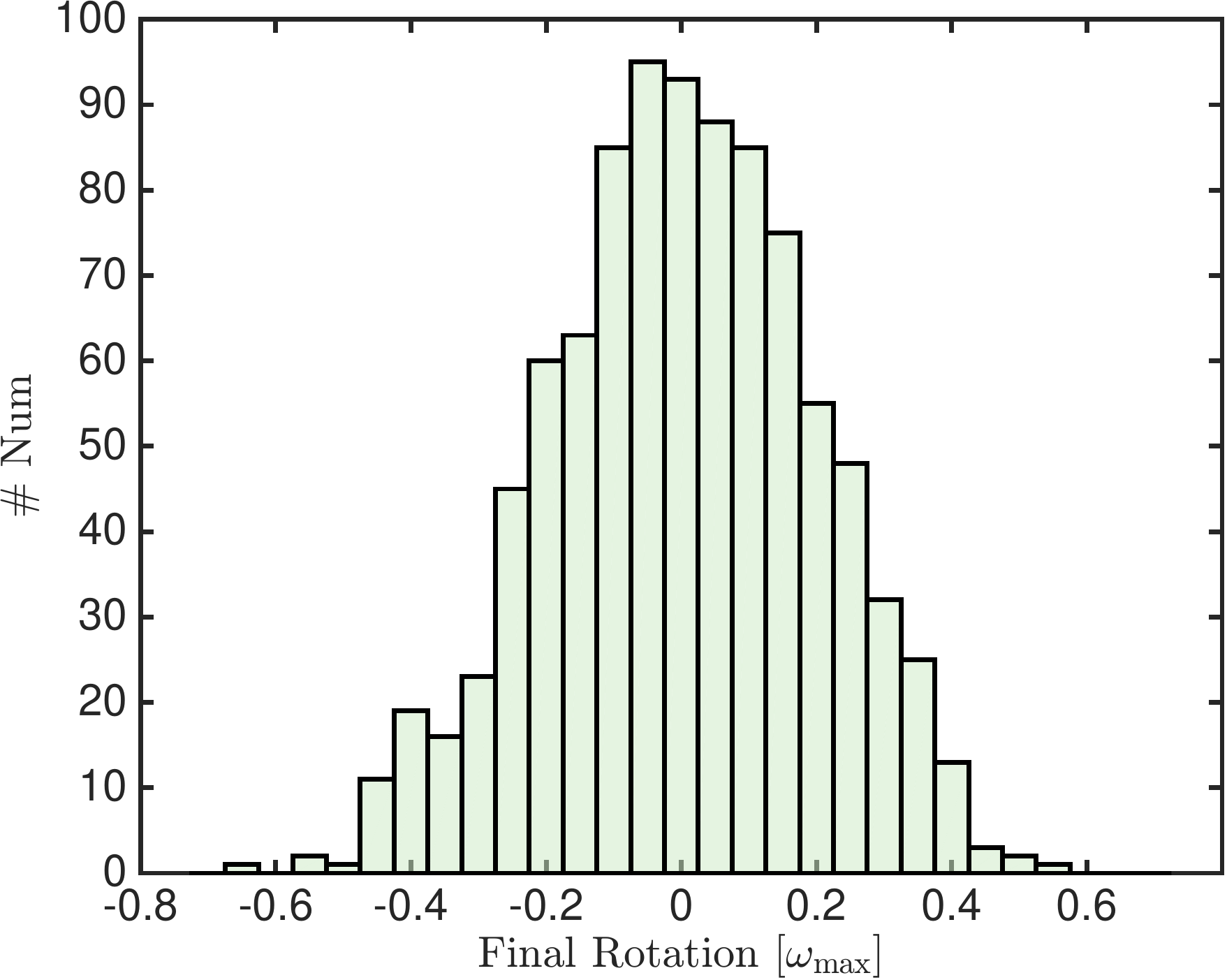}
\par\end{centering}
\end{singlespace}
\centering{} \renewcommand\figurename {Supplementary Figure} \caption{\textbf{Monte Carlo simulations.} Histogram of a) final moon mass;
b) the compositional difference between the final moon and the target.
Color bars represent different number of impact sequences. The red
dashed line represents the Earth-Moon oxygen isotope upper limit difference
(Young2016). c) The final rotation of the planet assuming sequence
of 30 impacts. }
\end{figure}

\section{Supplemental Movie Caption}

Movie 1 : Example of formation of a sub-lunar disk. The impact uses
a $\gamma=0.025$ impactor with a velocity of $2V_{{\rm esc}}$, at
an impact angle of $\beta=30^{o}$ and a fast initial planetary rotation
($0.5\,\omega_{{\rm max}}$). The color bars represent the entropy
of the impactor and target. The movie captures the equatorial plane
cross-section and the two parts overlap for display reasons. To emphasis
the merging of the cores, the impactor core is shown over the target.
The Roche limit is represented by the dashed line.

\section{Supplemental Result Table Caption}

Table 1: Impact outcomes. $\omega$ - initial spin of the target,
normalized by the break-up rotation \foreignlanguage{english}{$\omega_{{\rm max}}$}.
$\gamma$ - impactor mass ratio $M_{{\rm imp}}/M_{{\rm tot}}$. $V_{{\rm imp}}$
- velocity of the impactor in ${\rm cm}\ s^{-1}.$ $\beta$ impact
angle in degrees, negative values indicate retrograde impacts as compared
to the planet spin vector. $L_{planet_{Initial}}$ the initial angular
momentum of the planet $[L_{{\rm EM}}]$. $L_{{\rm imp}}$ - impactor
angular momentum $[L_{{\rm EM}}]$. $M_{{\rm disk}}$ - mass of the
disk $[M_{{\rm moon}}]$. $M_{{\rm esc}}$ - escaped mass $[M_{{\rm moon}}]$.
$M_{{\rm planet}}$ - final planet mass $[M_{\oplus}]$. $M_{{\rm sat}}$
- mass of the satellite $[M_{{\rm moon}}]$. $L_{{\rm planet}}$ -
final planetary angular momentum $[L_{{\rm EM}}]$. $L_{{\rm disk}}$
- final disk angular momentum $[L_{{\rm EM}}]$. Iron in disk - percentage
of iron material in disk. Imp in disk - percentage of impactor material
in disk.

\selectlanguage{english}%

\selectlanguage{american}%
\bibliographystyle{naturemag}